\documentclass[usenatbib]{mn2e}

\usepackage{times}
\usepackage{graphicx}
\title[Chaos around black holes with discs or rings]
      {Free motion around black holes with discs or rings:\\
       between integrability and chaos --- II}
\author[O. Semer\'ak and P. Sukov\'a]
       {O. Semer\'ak
        and
        P. Sukov\'a\thanks{E-mail: lvicekps@seznam.cz}\\
       Institute of Theoretical Physics,
       Faculty of Mathematics and Physics,
       Charles University in Prague,
       Czechia}
\begin{document}

\date{}

\pagerange{\pageref{firstpage}--\pageref{lastpage}} \pubyear{}

\maketitle

\label{firstpage}

\begin{abstract}
We continue the study of time-like geodesic dynamics in exact static, axially and reflection symmetric space-times describing the fields of a Schwarzschild black hole surrounded by thin discs or rings. In the previous paper, the rise (and decline) of geodesic chaos with ring/disc mass and position and with test particle energy was revealed on Poincar\'e sections, on time series of position or velocity and their power spectra, and on time evolution of the orbital `latitudinal action'. In agreement with the KAM theory of nearly integrable dynamical systems and with the results observed in similar gravitational systems in the literature, we found orbits of very different degrees of chaoticity in the phase space of perturbed fields. Here we compare selected orbits in more detail and try to classify them according to the characteristics of the corresponding phase-variable time series, mainly according to the shape of the time-series power spectra, and also applying two recurrence methods: the method of `average directional vectors', which traces the directions in which the trajectory (recurrently) passes through a chosen phase-space cell, and the `recurrence-matrix' method, which consists of statistics over the recurrences themselves. All the methods proved simple and powerful, while it is interesting to observe how they differ in sensitivity to certain types of behaviour.
\end{abstract}

\begin{keywords}
gravitation -- relativity -- black hole physics -- chaos
\end{keywords}

\section{Introduction}

The black holes discussed in most university courses are isolated, stationary and living in asymptotically flat space-times, that is to say, they belong to the Kerr(-Newman) family. The metric which describes this family is not entirely simple, but it has a number of ``nice" properties. Among others, its multipole structure---namely that required for the black-hole uniqueness theorems to work---is just the one that permits the solution of (electro-)geodesic equations in terms of separated first integrals (e.g. \citealt{Will-09}). This full integrability is mostly lost if any of the assumptions (isolation, stationarity, asymptotic flatness) is released.
More precisely, the integrability holds in the whole family of Kerr-Newman-NUT-(anti-)de Sitter space-times \citep{Carter-68-CMP}; it is connected with the existence of an irreducible second-order Killing tensor \citep{WalkerP-70} and has been shown to follow from the existence of a principal (conformal) Killing-Yano tensor there (e.g. \citealt{FrolovK-08}). All these space-times are of Petrov type D and represent subclass of the Pleba\'{n}ski-Demia\'{n}ski solutions with non-accelerated sources. Besides mass, electric charge and rotational angular momentum, they also include cosmological constant, magnetic charge and NUT parameter (of which the last two do not seem to be physically relevant, however).

Although the Kerr(-Newman) metric is being referred to when speaking about black holes in galactic nuclei and X-ray binaries, the above assumptions can only be valid approximately in such astrophysical circumstances; strictly speaking, they are all violated. Indeed, the observability of the supposed black holes alone implies that they have to be interacting with matter, thus non-isolated and non-stationary. Black holes certainly dominate the gravitational potential and intensity in their wide surroundings, but higher derivatives of the field (curvature) may be affected by nearby matter significantly. And these higher derivatives govern stability of motion. Hence, due to its own gravity, the matter may in fact settle down, around a central black hole, to a different configuration than which would be assumed by a {\em test} (non-gravitating) matter.

In a non-linear theory like general relativity it is not simple to specify what violation of the given (Kerr) metric is already large enough to invalidate various related conclusions and to bring physical differences with observable consequences. If the ambient matter is dilute and its self-gravitation effects are correspondingly weak, or when the source is more concentrated but only the field farther from it is relevant, the problem is usually being addressed by perturbation techniques. However, perturbation solutions are given in terms of series that practically must be truncated somewhere, so they cannot represent fully the self-gravitation effects (and in linear order they do not encompass them at all). The perturbative description is mainly questionable in the case of two- or one-dimensional additional sources (like discs or rings), because even if the total mass of such sources is very small, they however constitute a singularity and thus cannot be considered weak in their vicinity.

It is sure in any case that almost any deviation from the Kerr(-Newman) metric implied by the presence of additional matter leads to the loss of complete (electro-)geodesic integrability. This is even true when the additional source keeps the symmetries of the ``original", pure black-hole space-time, i.e. when their resulting ``superposition" is stationary, axially and reflection symmetric and orthogonally transitive (this last property is ensured if and only if the source elements do not perform any other motion than steady orbiting along the direction of axial symmetry). The loss of integrability in turn means that motion in the field of such a system is chaotic in general. The robustness of this conclusion makes the subtle phenomenon of chaos one of the aspects of astrophysical black-hole systems that should be taken into account and that can have observable consequences (see e.g. \citealt{Lukes-GAC-10}).

Needless to say, in astrophysical systems with accreting black holes the matter elements have many other (and more serious) reasons why to behave in a chaotic way. They extend from micro-scale processes over (magneto)hydrodynamics of accretion to interaction with (chaotic) radiation. However, as opposed to the case of discrete sources moving in the field of accreting stellar-mass black hole (e.g. in an X-ray binary), these ``physical" reasons for chaos should be less important in the case of whole stars orbiting a supermassive black hole in a galactic nucleus. Under the presence of a heavy accretion disc (and/or massive gas torus farther away), the motion of such ``test particles" may exhibit chaotic features on a sufficient time-scale, due to the ``gravitational" reasons alone. Admittedly, there are whole star clusters rather than single stars around the nuclear black holes, so each individual star would also ``feel" perturbations from all the other stars. Such a many-body problem is very difficult to tackle in general relativity and even in Newtonian theory it is being treated numerically. A possible simplification is to approximate the influence of the whole cluster on one particular star by means of an additional spherical or other simple-shape potential (see e.g. \citealt{KarasS-07,MadiganLH-09,LockmannBK-09}).

In spite of the ``shadowing theorem" \citep{Palmer-00}, it often seems hopeless to model the behaviour of a realistic non-linear dynamical system in detail, because sensitive dependence on initial conditions---one of the characteristic signs of chaos---involves sensitive dependence of conclusions on our ability to describe and treat the system {\em exactly} (see e.g. \citealt{JuddS-09}). Unfortunately, the non-linearity of general relativity as the theory of the underlying configuration space adds another piece of ``sensitivity" to the problem, and furthermore it strongly limits our compass to treat the problem exactly. In particular, the non-linearity severely limits the possibility of describing exactly the gravitational field (i.e. the configuration space) of multi-component systems, even if they are not extended. At present, the exact analytic treatment of black holes with additional sources is only practically possible in static and axially symmetric case (thus even rotation is excluded in general) with zero cosmological constant. Outside of the sources, the complete space-time can then be described by the (Weyl) metric containing just two unknown functions, one of which has the meaning of Newtonian gravitational potential. In a vacuum, the Einstein equations yield Laplace equation for this potential (like in the Newtonian description), so its ``total" form is obtained simply by adding the contributions from individual sources. The ``non-Newtonian" part of the problem lies in finding the second unknown metric function by a line integral; it is rather an exception than a rule that this can be accomplished explicitly.

We will however not repeat this standard introduction to the static and axially symmetric problem; it was summarised (e.g.) in the first paper \citep{SemerakS-10}. There, we placed uncharged annular thin discs without radial pressure (and without heat transfer) or their limit---(one-dimensional) rings---symmetrically about the (originally Schwarzschild) black hole in order to approximate the configuration of matter assumed in black-hole accretion systems. Considering, in particular, several discs of the inverted Morgan-Morgan counter-rotating family \citep{LemosL-94,Semerak-03} and the Bach-Weyl ring \citep{BachW-22,D'AfonsecaLO-05} as the additional sources, we studied how the dynamics of time-like geodesics in the field of such a system depends on parameters, mainly on relative mass and position of the external disc/ring and on the energy of test particles. The system showed typical features of a weakly non-integrable dynamical system. We observed, on Poincar\'e sections and on time series of phase variables and their power spectra, how it gradually turns chaotic when relative mass of the disc/ring or energy of the orbits are increased; we also noticed, quite generically, that for very high values of these parameters the system rather recurred to more regular behaviour.

It is a conventional experience how the originally fully regular phase space grains into chains of resonance islands, circumscribed by separatrices from which a net of chaotic filaments originates that gradually spreads and fuses into a ``chaotic sea". However, besides the overall development of the phase space, it is also interesting to study individual trajectories and try to distinguish different types among them. Actually, it is known (KAM theorem) that in ``weakly perturbed" systems, the phase space contains regions of very different degrees of chaoticity for almost any ``strength" of the perturbation agent. Moreover, even a given {\em single} orbit may show different degrees of chaoticity/regularity within its different stages. This is especially valid for the orbits prone to ``sticky motion", i.e. those which spend a long time very close to regular islands, while only occasionally diverging into a chaotic sea. We already singled out several such orbits in the first paper \citep{SemerakS-10} and illustrated that they can be distinguished from ``strongly chaotic" orbits, drowned in the chaotic sea, according to the power spectra of phase-variable time series (that of ``vertical" position, $z=z(t)$, for example). In general relativity, this simple method was notably employed by \citet{KoyamaKK-07} who demonstrated (on the problem of spinning particles in a Schwarzschild background) that the power spectra of ``strongly chaotic" orbits have ``white-noise" low-frequency part (relatively flat curve at rather low values), whereas the spectra of ``weakly chaotic" orbits incline to the ``1/frequency" shape at low frequencies (and rise to higher values there). This is natural since stronger chaos means more irregular time series with less distinct periods, in particular without marked recurrences even on longer timescales.

In the present paper, we analyse in more detail several individual orbits selected out of those which---at least for a certain time---adhere to regular regions (``sticky motion"). We show on Poincar\'e diagrams and on power spectra of the corresponding ($z$-position) time series that different parts of these orbits present different degree of chaoticity. In stages when the orbits tend to fill uniformly a large area (chaotic sea), the spectrum approaches white noise at low frequencies, whereas almost regular parts of the same orbits indeed generate ``1/frequency" shape there. However, even ``highly regular" parts of such world-lines can be clearly distinguished from strictly regular orbits. On the other hand, it is known that even strongly chaotic {\em deterministic} behaviour is clearly distinguishable from a random one. It is our object here to check the above for selected orbits of our dynamical system. We compare how the character of geodesic dynamics is revealed by $z$-position power spectra (section \ref{spectra}), by the ``average directional vectors" method based on statistics over the directions in which the trajectory (recurrently) passes through specified phase-space cells (section \ref{directions}; see \citealt{Sukova-11} for preliminary results), and by ``recurrence plots" which visualise the pattern of recurrences to such cells (section \ref{recurrences}). Several useful quantifiers of chaos following from the recurrence plots will also be computed and plotted.

\section{Power spectra of geodesics and of their parts}
\label{spectra}

\begin{figure*}
\includegraphics[width=\textwidth]{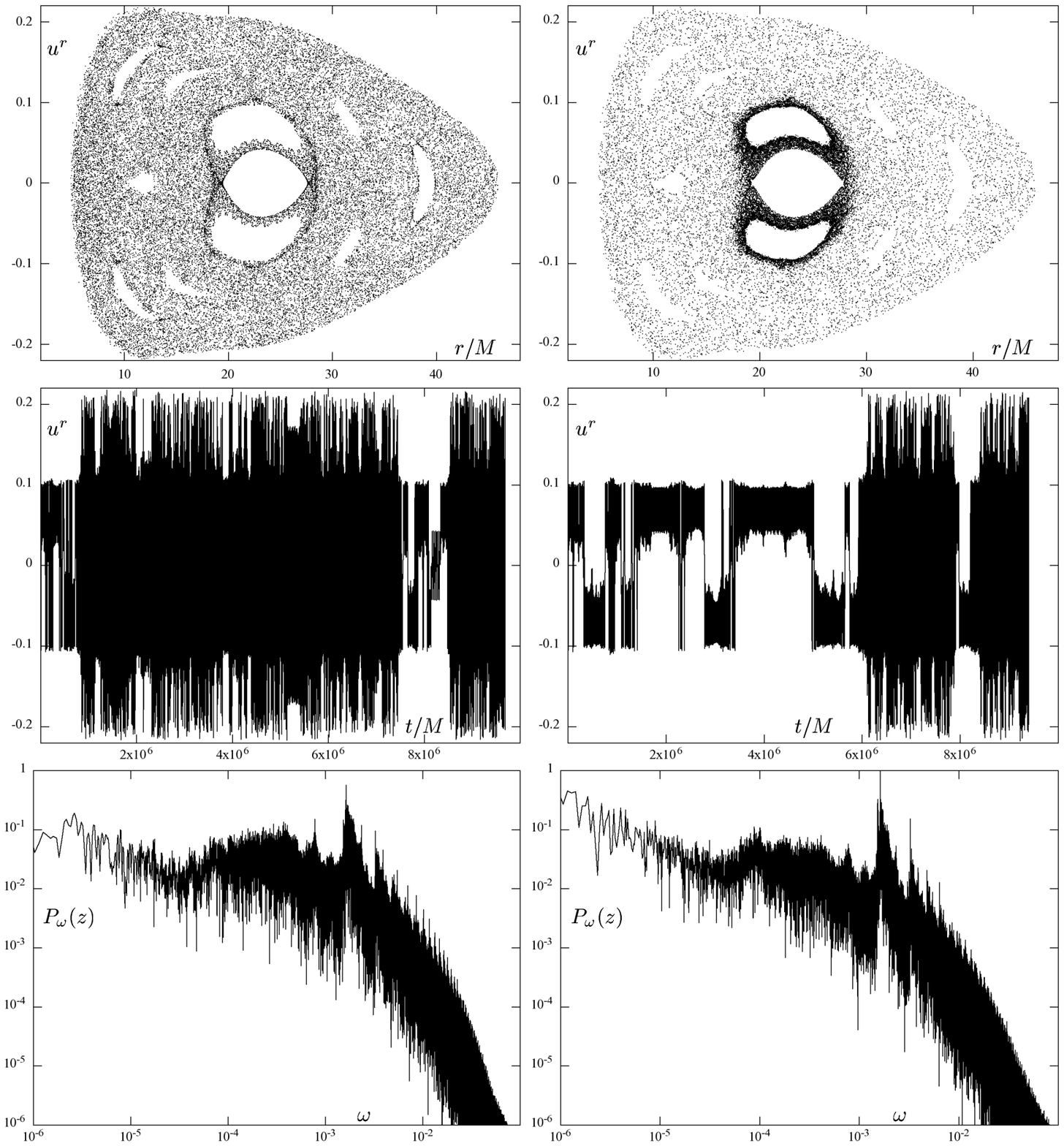}
\caption
{Two geodesic orbits in the field of a black hole (of mass $M$) surrounded by the inverted first Morgan-Morgan disc with mass ${\cal M}=1.3M$ and inner radius $r_{\rm disc}=20M$. Both geodesics have specific energy and specific angular momentum at infinity ${\cal E}=0.956$ and $\ell=4M$. The orbit on the left is clearly more chaotic, the one on the right spends more time in the vicinity of central regular islands in the Poincar\'e diagram (showing passages across the equatorial plane, top row). The time series of $u^r$ (given by values recorded at passages through the equatorial plane) are plotted in the middle row and the power spectra of the $z(t)$ evolution are plotted in the bottom row.}
\label{orbits-whole}
\end{figure*}

\begin{figure*}
\includegraphics[width=\textwidth]{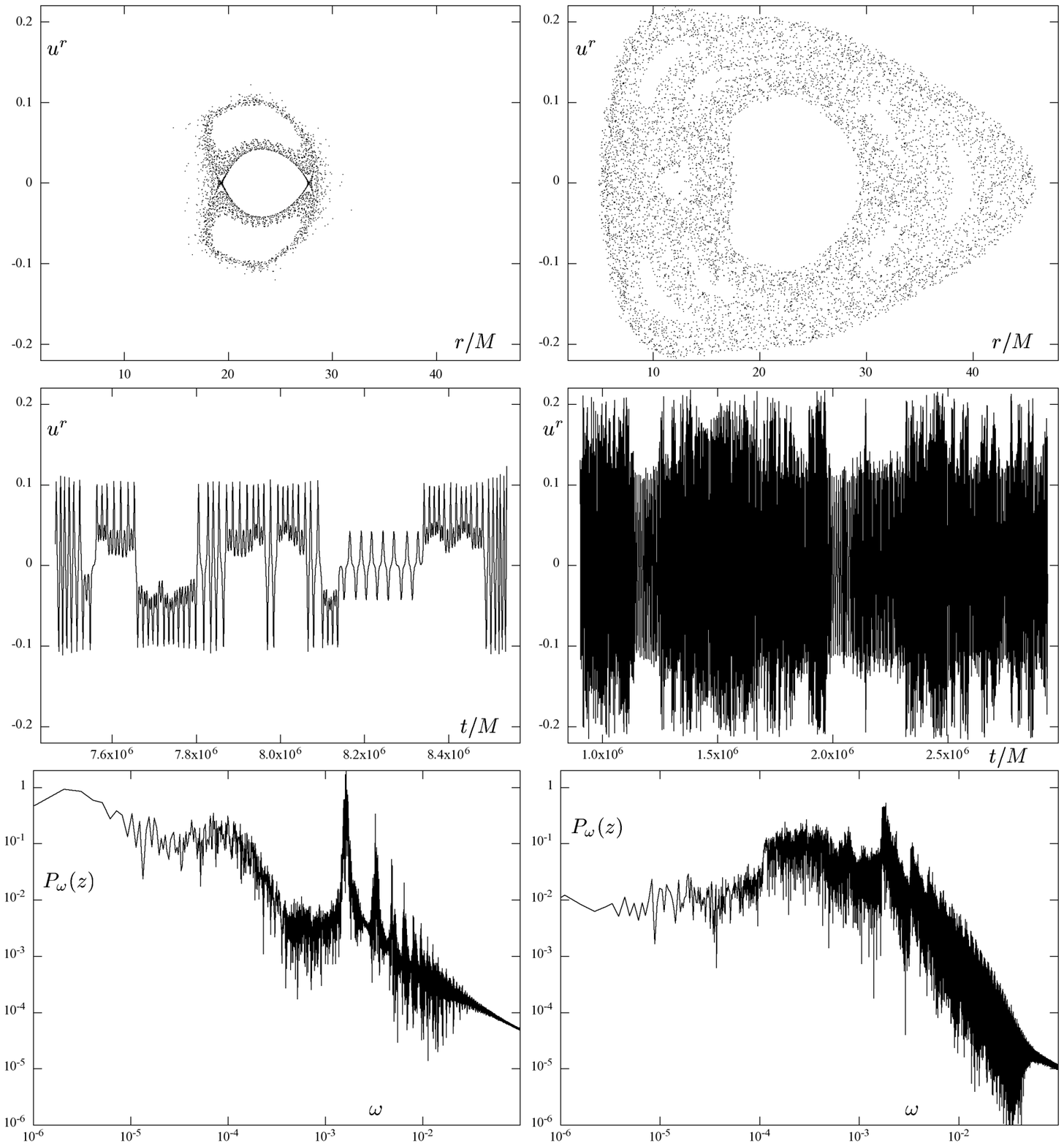}
\caption
{Example of a rather regular (left column) and rather chaotic (right column) phases of the orbit shown in the left column of figure \ref{orbits-whole}. Equatorial Poincar\'e section (top), $u^r(t)$ time series recorded at passages through the equatorial plane (middle) and power spectrum of the $z(t)$ evolution (bottom) are shown.}
\label{orbit0-excerpts}
\end{figure*}

\begin{figure*}
\includegraphics[width=\textwidth]{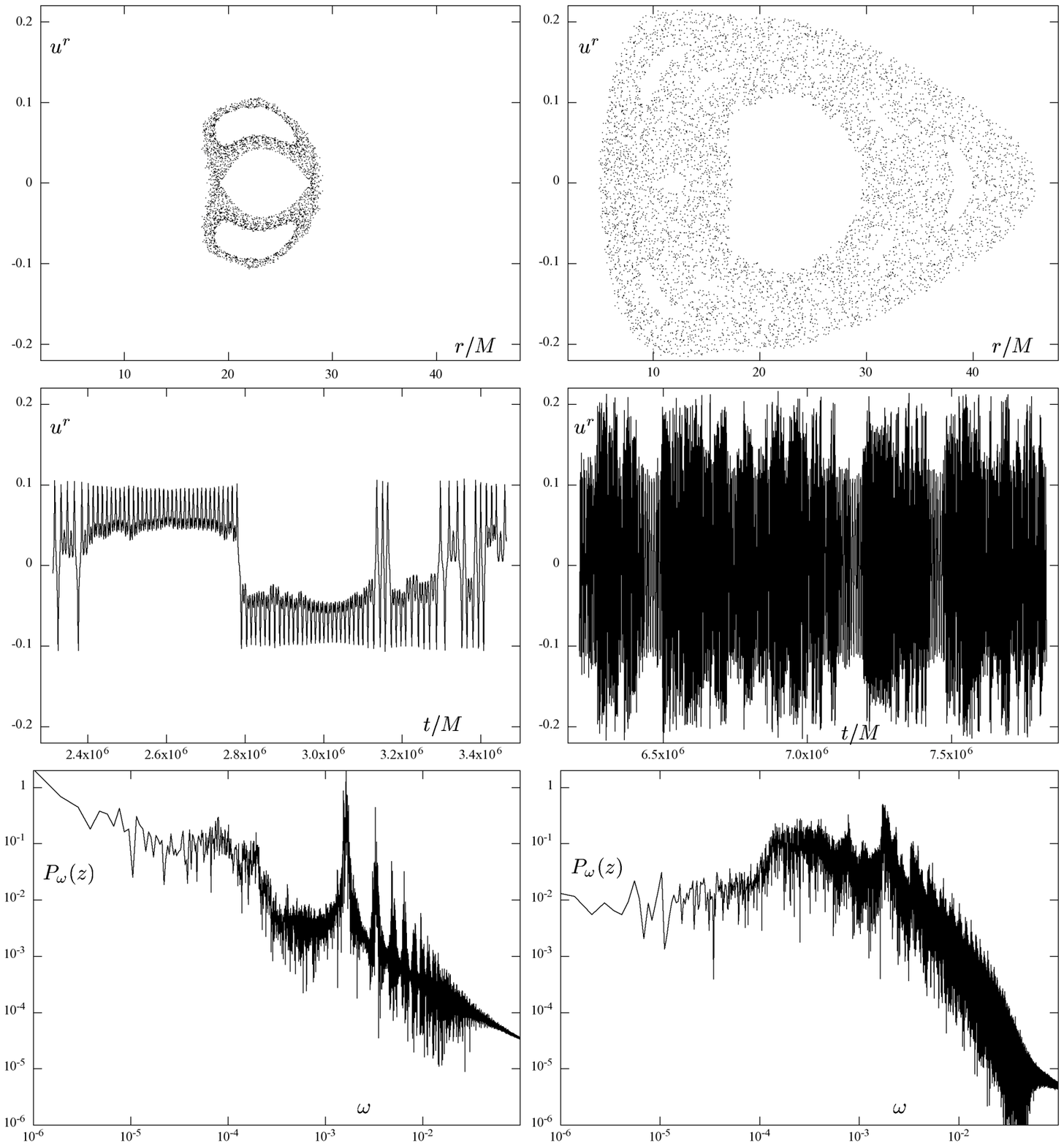}
\caption
{Example of a rather regular (left column) and rather chaotic (right column) phases of the orbit shown in the right column of figure \ref{orbits-whole}. Equatorial Poincar\'e section (top), $u^r(t)$ time series recorded at passages through the equatorial plane (middle) and power spectrum of the $z(t)$ evolution (bottom) are shown.}
\label{orbit1-excerpts}
\end{figure*}

\begin{figure*}
\includegraphics[width=\textwidth]{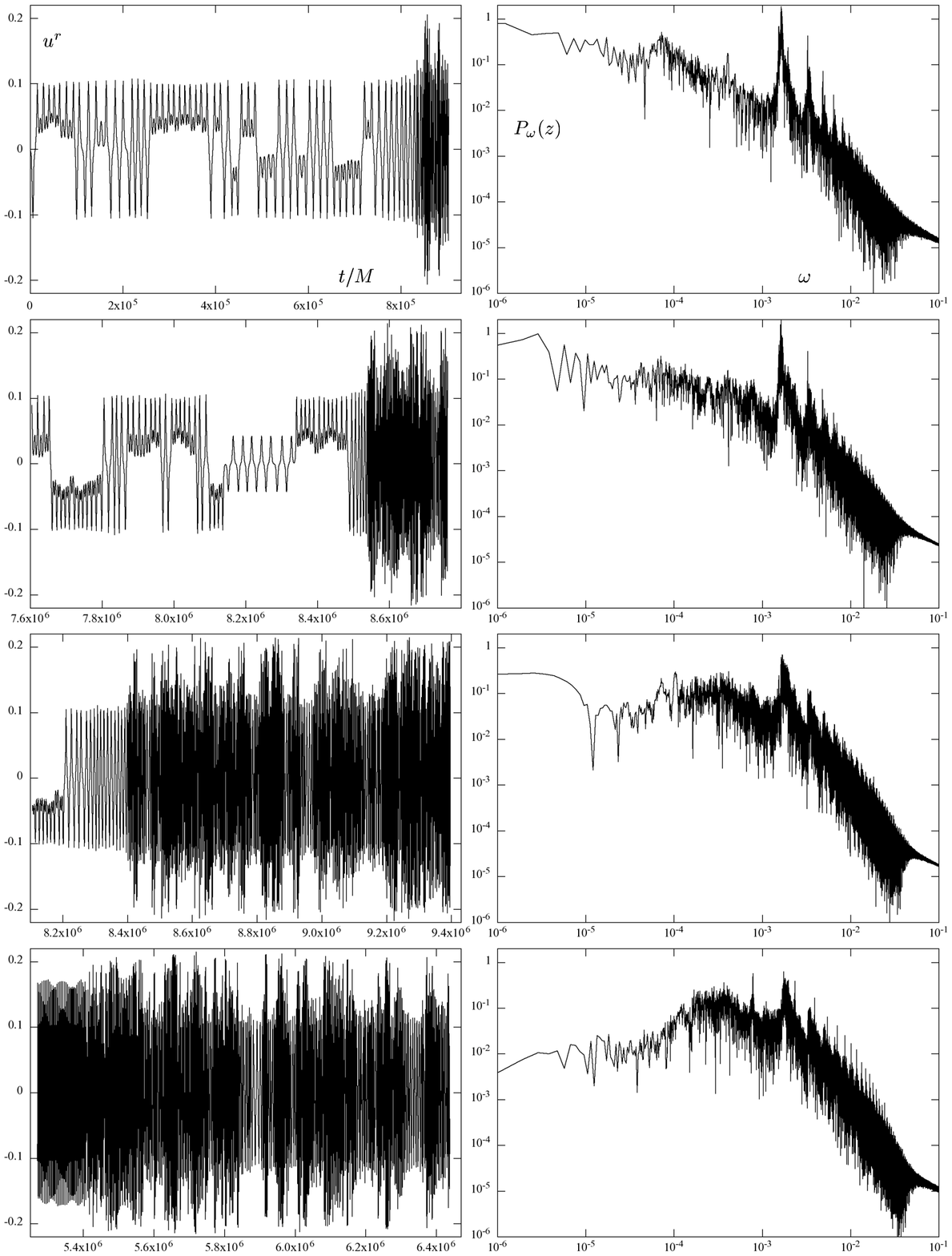}
\caption
{The $u^r(t)$ behaviour (left column) and power spectra of $z(t)$ evolution (right column) of four different sections of orbits shown in figure \ref{orbits-whole}; taken from top to bottom, the 1st, the 2nd and the 4th sections are parts of the 1st orbit (shown in the left part of figure \ref{orbits-whole}), while the 3rd section belongs to the 2nd orbit (right part of figure \ref{orbits-whole}). The degree of chaoticity clearly grows from top to bottom.}
\label{series}
\end{figure*}

Poincar\'e surfaces of section give a picture of how much regular/chaotic the system with given parameters is. It is a property of ``nearly-integrable systems" that some quite irregular trajectories already appear after a tiny perturbation, and vice versa, even a strongly perturbed phase space still harbours some quite regular ones. It can also be recognised on Poincar\'e diagrams that (in rather strongly perturbed cases) certain orbits behave quite differently within different periods, namely they alternately stick to islands of regular motion and drift around the ``chaotic sea". Hence, the ``degree of chaoticity" can be ascribed to the system (phase space) on the whole, but also to its individual orbits, while in a sense it is a property of a given part of a given orbit. It should be stressed, however, that such a tracking of regular/chaotic features down to a particular segments of particular orbits has to be taken with much caution, since it is in fact inconsistent with the essence of deterministic chaos as a {\em global} phenomenon. This is especially clear on systems like billiards where interaction only occurs at discrete events: one cannot say that their trajectories are almost all regular, with chaos solely occurring at this or that point. In systems where the interaction is long-range and acts continuously (and, ideally, even smoothly), like the gravitational one in our case, such attempts to ``localise" the character of dynamics are more propitious, but still they do not seem to yield necessary or sufficient criteria (cf. \citealt{VieiraL-96b}). The global nature of chaos go together well with the similar attribute of Fourier transform, which probably helps power spectra to be an efficient tool of study of dynamical systems.

We select two geodesic orbits in the field of a black hole (of mass $M$) surrounded by the inverted first Morgan-Morgan disc with mass ${\cal M}=1.3M$ and inner radius $r_{\rm disc}=20M$. Such a disc is clearly very massive in comparison with what is considered astrophysically realistic; it causes quite a strong perturbation of the original black-hole field and the geodesic dynamics is rather chaotic in general (see paper I). The selected two orbits have specific energy and angular momentum at infinity ${\cal E}\equiv u_t=0.956$ and $\ell\equiv u_\phi=4M$, respectively,\footnote
{We use geometrised units in which $c=G=1$. The particle's proper time is denoted by $\tau$ and its four-velocity by $u^\mu$. The Schwarzschild-type coordinates $(t,r,\theta,\phi)$ are employed, with vertical position $z=r\cos\theta$. In particular, the time $t$ is tied to the time-like Killing symmetry of space-time.}
and initial conditions very close to those of the orbit drawn in light blue in figure 16 of the previous paper \citep{SemerakS-10}; the power spectrum of its $z$-position is clearly seen in the bottom right panel there, among several other ``weakly chaotic" orbits showing the ``1/frequency" spectral shape.
Note that the power spectrum is obtained using the discrete Fourier transform
\begin{equation}
  P_\omega(z)=\frac{1}{N}\,
              \left|\sum\limits_{n=0}^{N-1}
                    z(\tau_n) e^{-(2\pi/N){\rm i}\omega\tau_n}\right| \,,
\end{equation}
where $N$ is the total number of samples. The spectrum thus illustrates which frequencies dominate the particle motion perpendicular to the plane where the disc or the ring is placed.
Both selected orbits were followed for a long time (of about $10^7 M$ of proper time, which means for some $10^4\div 10^5$ orbital periods) and then divided in parts in order to check how these sections appear in Poincar\'e diagram and in the power-spectrum plot. Rather than showing the figures obtained for all the parts, let us just give several examples. In general, one can repeat the conclusion of \cite{KoyamaKK-07}, also supported in our previous paper: the orbits which appear ``strongly chaotic" in Poincar\'e diagram (filling rather uniformly thick layers there) produce irregular time series whose power spectra have white-noise character at low frequencies (rather flat at low level, without distinct peaks), whereas the orbits which appear ``weakly chaotic" in Poincar\'e diagram (remaining close to regular islands for considerable periods) produce almost regular time series in the periods of ``sticky motion" (while irregular elsewhere), which brings more power into low frequencies and the spectrum assumes $1/f$ shape there.

In figure \ref{orbits-whole}, the two selected orbits are represented as a whole, first on Poincar\'e sections showing passages through the equatorial plane in $(r,u^r)$ axes, then on the $u^r(t)$ behaviour and finally on the power spectra of $z(t)$ evolution. The Poincar\'e sections show that the first (left) trajectory spends some time in the vicinity of the split primary island, but generally it is rather chaotic. The second (right) trajectory fills the chaotic sea less densely and apparently prefers to stay in the layers adjacent to the central regular regions. The time series $z(t)$, $u^r(t)$ of the second trajectory really contain longer periods of almost regular oscillations, and also the power spectra of $z(t)$ confirm this difference in the above described sense: the second (right) spectrum is a bit less ``concave" than the first (left) one, it tends more to the power-law, $1/f$ shape, mainly at low frequencies (where it also ends at somewhat higher values). However, both trajectories contain rather chaotic as well as rather regular phases; we give two examples for each of the orbits in figures \ref{orbit0-excerpts} and \ref{orbit1-excerpts}. The sections shown on the left are confined to the vicinity of regular islands for considerable intervals, whereas those on the right are quite chaotic as is clear from the respective Poincar\'e diagrams, $u^r(t)$ courses and power spectra of $z(t)$ evolutions.

Several more specific features can be recognised in the figures. Comparing the weakly chaotic parts of the orbits, one observes that their rather ``$1/f$-shaped" power spectra can differ in slope and in noise degree, mainly in the high-frequency part. (Sure: it is important, among others, what is the periodicity of the island that the orbit adheres to.) Some of these spectra can really be almost fitted by a straight line; nevertheless, they typically have distinct peak in the middle part and a ``valley" to the left of this maximum. On the other hand, strongly chaotic parts of the orbits provide ``cat-back", concave spectral shape which cannot be approximated by a straight line; the spectra contain less distinct features, in particular, less distinct maximum and valley on its left. However, it would be misleading to generalise {\em any} tendency seen on certain particular series of spectra, because mainly the quasi-regular phases of motion can involve different amplitudes and periodicities and thus bring different spectral features; this is already clear from the time series themselves. In spite of it, we try to put together a series of orbital sections with different content of chaos/regularity and illustrate how it proves in power spectra (figure \ref{series}).

\section{Kaplan \& Glass' method of diagnosing the degree of chaos}
\label{directions}

\begin{figure*}
\includegraphics[width=\textwidth]{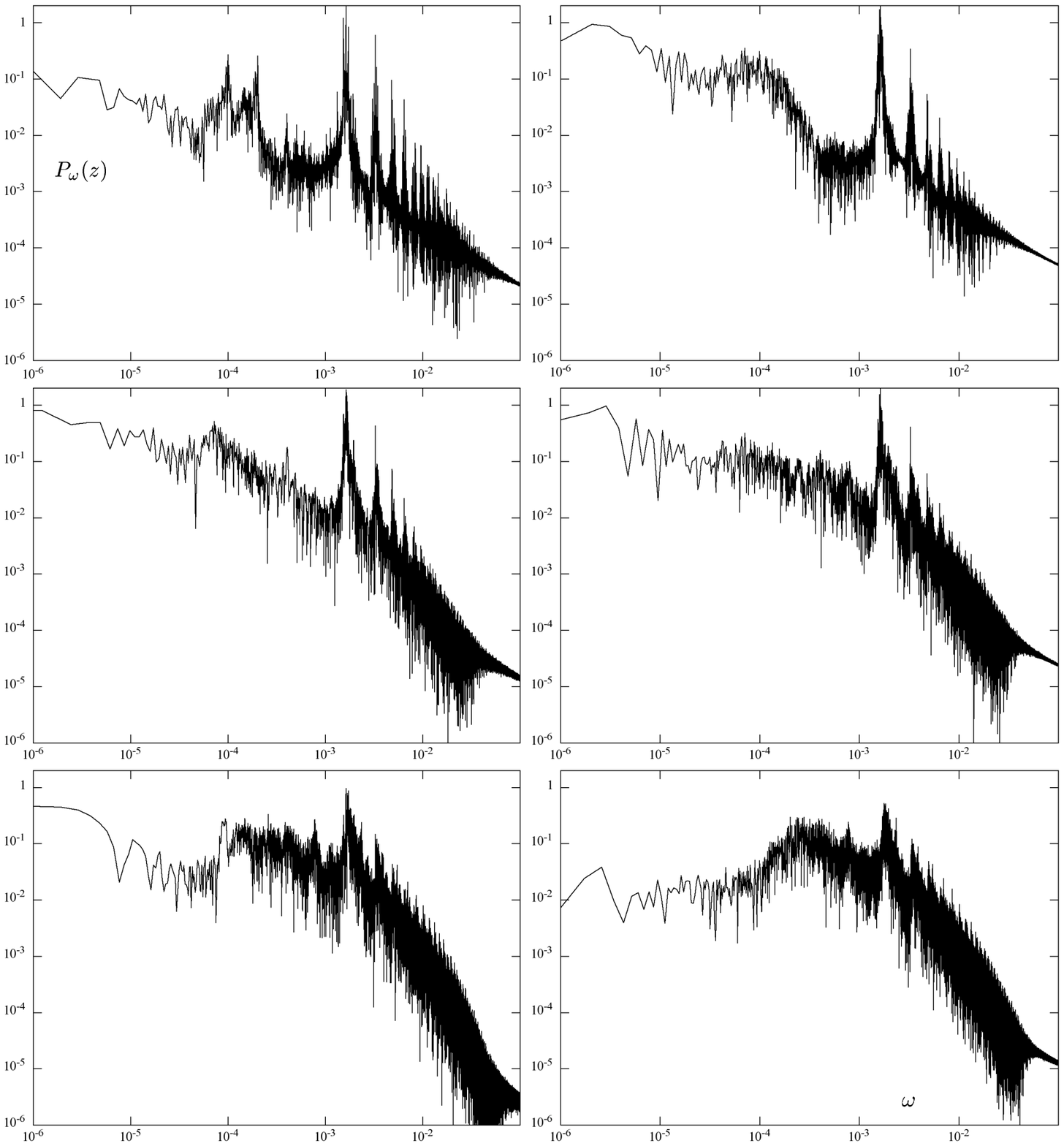}
\caption
{A: Power spectra of 6 different parts of the two orbits studied in the last section. In the order from top left to bottom right, their degree of chaoticity increases (but here on spectra it is barely obvious for the first 3 or 4 of them). The Kaplan-Glass function $\bar\Lambda(\Delta\tau)$ calculated for the same 6 orbital phases is shown in part B of this figure.}
\label{5-spectra}
\end{figure*}

\begin{figure*}
\includegraphics[width=\textwidth]{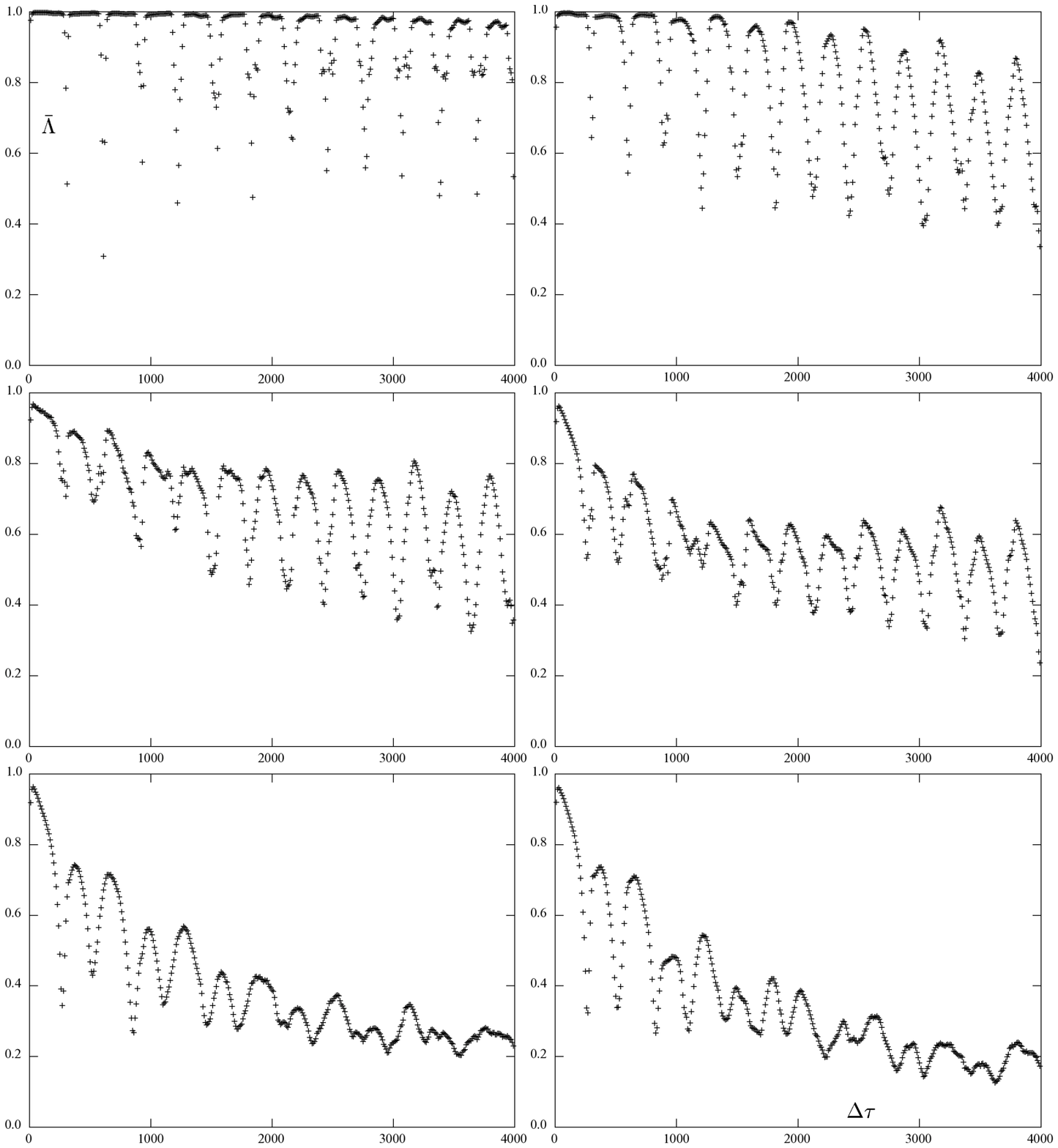}
\contcaption
{B: The Kaplan-Glass function $\bar\Lambda(\Delta\tau)$ calculated for the same 6 orbital phases whose power spectra are shown in previous figure \ref{5-spectra}.A. The behaviour of $\bar\Lambda(\Delta\tau)$ clearly distinguishes between these examples, in particular between the first four of them which appear quite similar according to spectra. On the other hand, the last two---quite chaotic---cases are similar here, although their power spectra show somewhat different degree of chaoticity.}
\end{figure*}

There exists a large number of methods how to recognise the degree of chaoticity---or even stochasticity---present in the time evolution of a given system. We will not give any comparative analysis, and not even a listing of them, but will just illustrate that processing of time series in some other way can yield information which is interesting to compare with what is revealed by power spectra. As an example of a simple but useful method, we will consider the one suggested by \cite{KaplanG-92}. It was designed to distinguish between deterministic and random systems, but we will see that it is quite sensitive and also able to recognise how much chaotic the (deterministic) system is.

The method of Kaplan \& Glass is based on monitoring the evolution of tangent to the trajectory in small subsets of phase space. However, the latter is not the ``original" phase space of the system (this may be unknown), but its $d$-dimensional embedding ``reconstructed" from a given data series $x(\tau)$ by taking the delayed replicas $x(\tau)$, $x(\tau-\Delta\tau)$, $x(\tau-2\Delta\tau)$, $\dots$, $x(\tau-d\Delta\tau)$ as its axes ($\Delta\tau$ is some real time shift); the reconstruction is justified by the delay embedding theorem of \citet{Takens-81}. Such a space is then coarse grained into a grid of $m^d$ cubes and average directions of passages through each of these cubes are added up. Namely, first the average direction of each ($k$-th) pass of a trajectory through the given ($j$-th) box is recorded as a unit form $\vec{v}_{kj}$ of the vector connecting the point where the trajectory entered the box with the point where it left it. Then the vectors obtained from a large number ($n_j$) of transits through the $j$-th box are summed (vector addition) and the length of the resulting vector $V_j$ is normalised by $n_j$,
\[V_j\equiv\left|\vec{V}_j\right|
     =\frac{1}{n_j}\sum\limits_{k=1}^{n_j}\vec{v}_{kj}\;.\]
Finally, the resulting norm $V_j$ is averaged over all boxes which were crossed $n$-times, and the dependence of this average ($\equiv\bar{L}^d_n$) on $n$ is checked. For random data, $\bar{L}^d_n$ decreases with $n$ roughly as $n^{-1/2}$; in particular, the average displacement per step for random walk in $d$ dimensions is (for large $n$) given by
\begin{equation}
  \bar{R}^d_n
  =\frac{\Gamma\!\left(\frac{d+1}{2}\right)}{\Gamma\!\left(\frac{d}{2}\right)}
   \;\sqrt\frac{2}{nd} \;,
\end{equation}
where $\Gamma(.)$ denotes the gamma function. (Note that $\bar{R}^{d\rightarrow\infty}_n=n^{-1/2}$.)
For a deterministic system, the average $\bar{L}^d_n$ decreases more slowly or even remains close to a maximal value of one; this is due to the fact that in every small neighbourhood the tangent vectors from all transits are almost parallel so the norm of their vector sum is almost maximal. (It depends on box size, however: theoretically, in the limit of infinitesimally fine grain, which is only conceivable for infinitely long data series, $\bar{L}^d_n=1$ for the deterministic dynamics.)

Besides the size of the lattice boxes, the result clearly depends on the dimension $d$ and the time lag $\Delta\tau$. In particular, the choice of $\Delta\tau$ may be a subtle issue, as discussed in the original paper \citep{KaplanG-92}. In order to analyse the dependence of $\bar{L}^d_n$ on $\Delta\tau$, one can compare the value of $V_j$ with $\bar{R}^d_{n_j}$ for each box and average this over all occupied boxes,
\begin{equation}
  \bar\Lambda(\Delta\tau)
  \equiv\left\langle
        \frac{(V_j)^2-(\bar{R}^d_{n_j})^2}{1-(\bar{R}^d_{n_j})^2}
        \right\rangle .
\end{equation}
In a theoretical limit, $\bar\Lambda=0$ for a random walk, whereas $\bar\Lambda=1$ for a deterministic system. In practice, with finite series and finite ``pixel" size, $\bar\Lambda$ falls off roughly as autocorrelation function for randomised signal, while more slowly for a deterministic signal.

For our system of geodesic dynamics in the static and axially symmetric field of a black hole surrounded by a disc or a ring, we choose $d=3$ and construct the ``phase space" out of the time series of the test-particle vertical position $z=r\cos\theta$ ($r$ and $\theta$ are Schwarzschild radius and latitude), i.e. the space is spanned by the axes $z(\tau)$, $z(\tau-\Delta\tau)$ and $z(\tau-2\Delta\tau)$. We choose its ``elementary-grain" size of the order of $M$ (thus volume $\simeq M^3$) and focus on the behaviour of $\bar{\Lambda}$ with increasing $\Delta\tau$. Computation of this dependence for a large number of orbits (or their parts) confirmed that the method of Kaplan \& Glass is able to reveal the degree of their chaoticity. For rather regular (``sticky") evolutions, $\bar{\Lambda}$ really remains very close to one except when $\Delta\tau$ is just multiple of some important orbital period. On the other hand, for strongly chaotic (parts of the) orbits, $\bar{\Lambda}$ has a peak almost approaching one for a certain value of $\Delta\tau$ (which is smaller than a period and is related to the first zero of the autocorrelation function) and than it decreases quite rapidly (with some oscillation), because around such orbits the geodesic flow is sensitive to initial conditions and the deterministic connection between the position $x(\tau)$ and $x(\tau-\Delta\tau)$ fades away quickly with the growing time lag. The rate of decrease of $\bar\Lambda(\Delta\tau)$ is a simpler (more easily comparable) indicator of the degree of chaoticity of the orbit than the power spectrum. Namely, it is just {\em value} of $\bar\Lambda(\Delta\tau)$ what is important, whereas the spectrum mainly bears its information in the overall shape and the ``degree of noisiness", of which especially the latter may be difficult to judge and compare. At the same time, the Kaplan-Glass $\bar\Lambda(\Delta\tau)$ indicator seems to be quite reliable, and also ``fine" in the sense that it can distinguish between evolutions which appear very similar (or just not easily comparable) on spectra.

Let us illustrate this on several parts of our two orbits discussed in previous section. Figure \ref{5-spectra} shows spectra of 6 different orbital sections, of which the 1st and the 5th are parts of the second orbit and the rest are parts of the first orbit. The sections are placed (top left, top right, middle left, etc.) in the order of increasing chaos. Figure \ref{5-spectra}.B shows the $\bar\Lambda(\Delta\tau)$ functions for the same 6 orbital sections. It is seen how the $\bar\Lambda(\Delta\tau)$ behaviour safely distinguishes between the first four examples which are all ``rather regular" and have just slightly different character of spectra. (But after the clue is provided by $\bar\Lambda(\Delta\tau)$, one admits that the spectra also reveal the same tendency.) On the other hand, $\bar\Lambda(\Delta\tau)$ appears not to be so sensitive in more chaotic regions: starting from a certain amount of chaoticity, it does not decrease any more and gives similar behaviour for orbital phases which still quite differ in spectra, mainly in the low-frequency tail. (For instance, the last two examples of figure \ref{series} yield the same course of $\bar\Lambda(\Delta\tau)$.)

\begin{figure*}
\includegraphics[width=\textwidth]{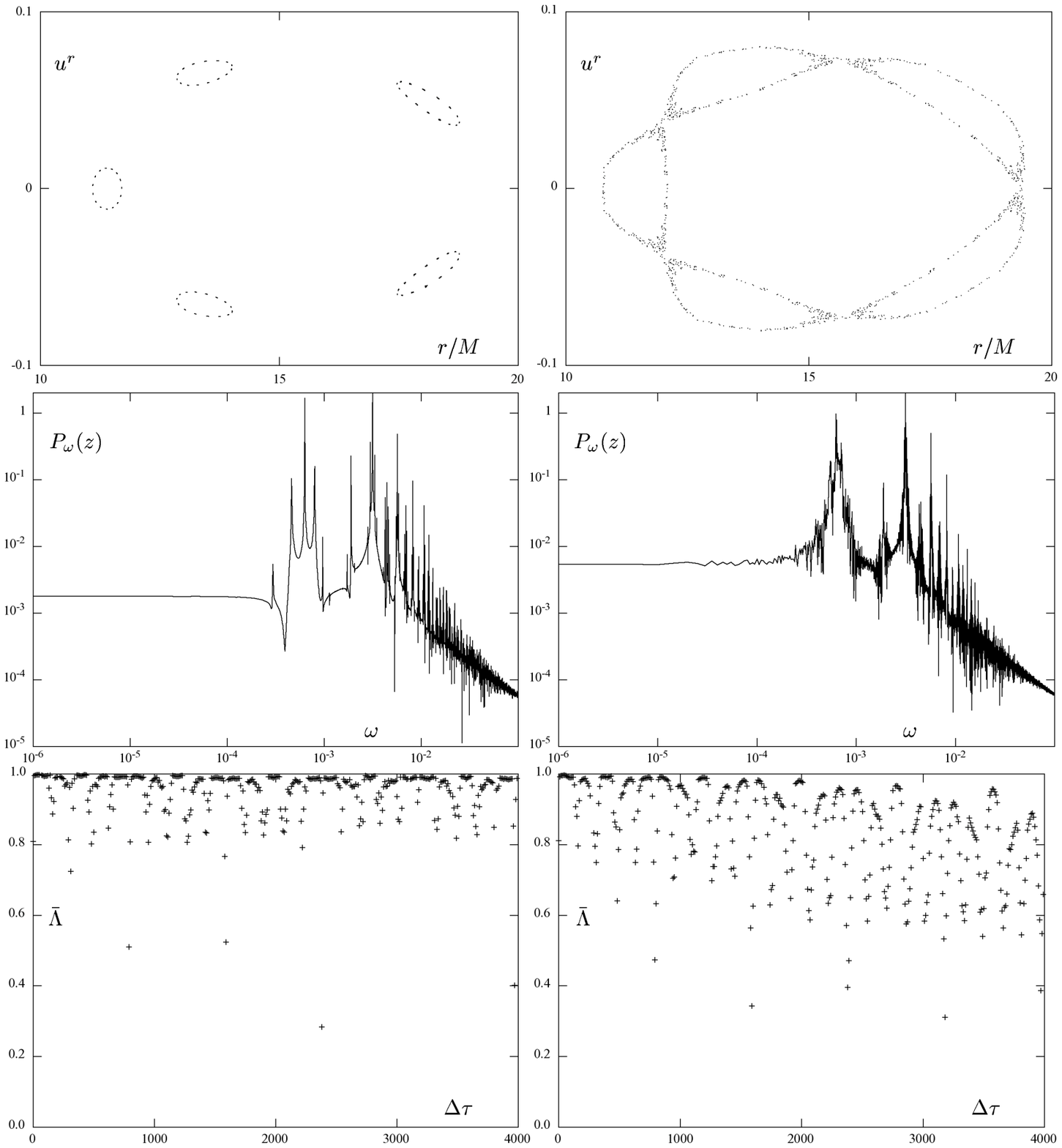}
\caption
{Two geodesic orbits in the field of a black hole (of mass $M$) surrounded by the inverted first Morgan-Morgan disc with mass ${\cal M}=0.5M$ and inner radius $r_{\rm disc}=14M$. Both geodesics have specific energy and specific angular momentum at infinity ${\cal E}=0.955$ and $\ell=3.75M$. The first (left column) belongs to five-periodic regular islands and the other (right column) sticks to the first rather closely on Poincar\'e diagrams (top row). The slight difference between the orbits also shows on power spectra of their $z(t)$ evolution (middle row) and on the behaviour of their tangent's autocorrelation represented by the $\bar\Lambda(\Delta\tau)$ dependence (bottom row).}
\label{iMM-other}
\end{figure*}

\begin{figure*}
\includegraphics[width=\textwidth]{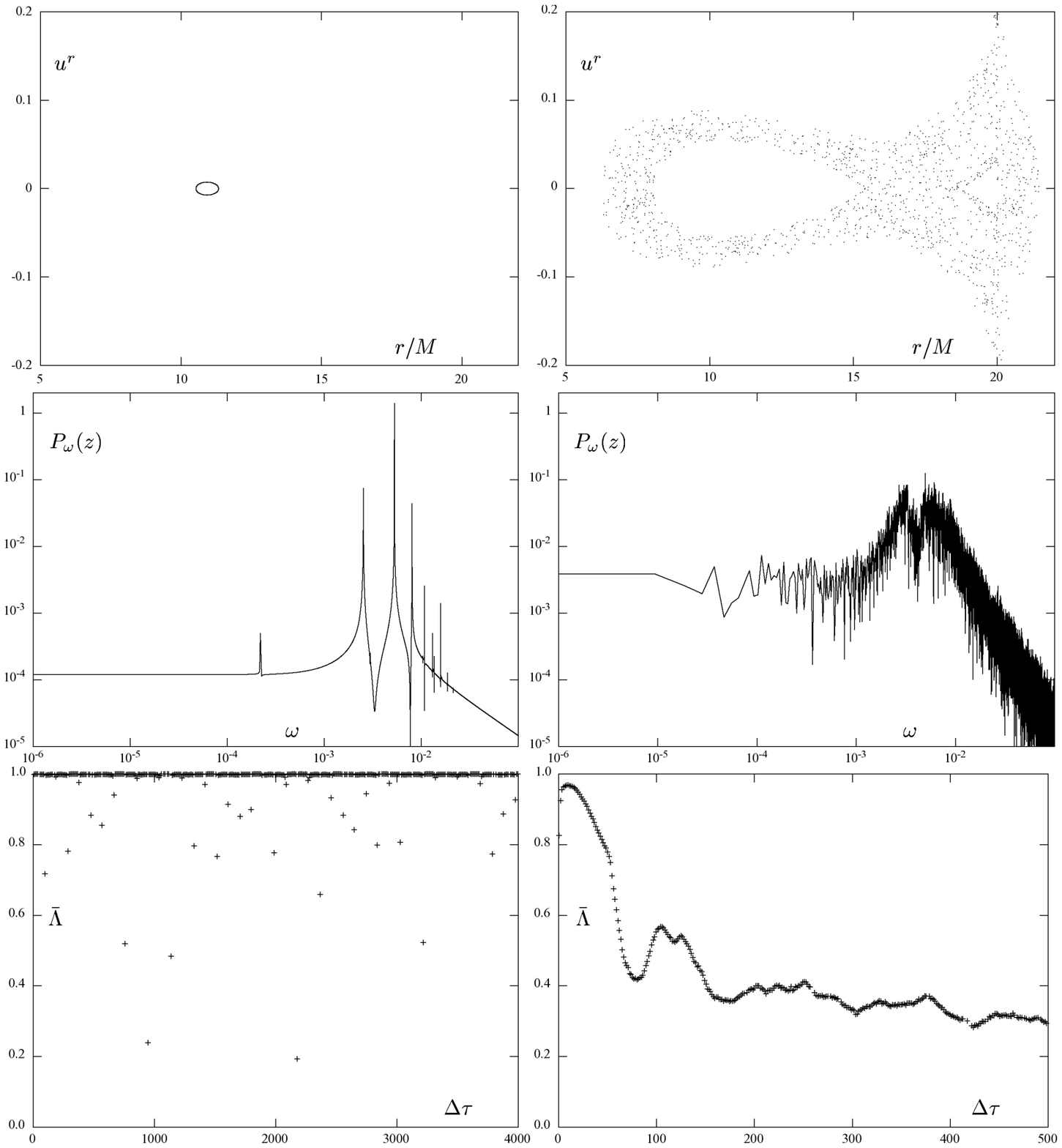}
\caption
{Two geodesic orbits in the field of a black hole (of mass $M$) surrounded by the Bach-Weyl thin ring with mass ${\cal M}=0.5M$ and radius $r_{\rm ring}=20M$. Both geodesics have specific energy and specific angular momentum at infinity ${\cal E}=0.93$ and $\ell=3.75M$. The first (left column) lies deep in the primary regular island, while the other (right column) fills the chaotic sea around and in the vicinity of the ring (see Poincar\'e diagrams in top row). The big difference between the orbits also shows on power spectra of their $z(t)$ evolution (middle row) and on the behaviour of the Kaplan-Glass averaged autocorrelation parameter $\bar\Lambda(\Delta\tau)$ (bottom row).}
\label{BW-example}
\end{figure*}

We add two more illustrations, one (figure \ref{iMM-other}) showing two orbits in the field of a black hole with the inverted Morgan-Morgan disc of different parameters than above (mass ${\cal M}=0.5M$ and inner radius $r_{\rm disc}=14M$), and the other (figure \ref{BW-example}) showing two orbits in the field of a black hole surrounded by a thin, Bach-Weyl ring (of mass ${\cal M}=0.5M$ and radius $r_{\rm ring}=20M$). The orbits in figure \ref{iMM-other} are only slightly different, one belonging to five-periodic regular islands and the other sticking closely to these islands; you can see how they differ in power spectra and in the $\bar\Lambda(\Delta\tau)$ dependence. The orbits presented in figure \ref{BW-example} differ much more, one lying very close to the central circular orbit of the primary regular island, while the other filling the chaotic sea around and in the vicinity of the ring.

\section{Recurrence analysis of the orbits}
\label{recurrences}

Let us stress again that the above method is just an example of a simple but apparently quite reliable way of judging and classifying the regularity / chaoticity / stochasticity of a given time series. Another example of a simple and powerful method is the recurrence analysis which is based on checking the recurrence of orbits of a dynamical system to a chosen (small) cell of phase space (see \citealt{MarwanCTK-07} for a thorough survey). The pattern of recurrences encodes quite credibly the character of the dynamics, clearly distinguishing between regular, chaotic and random evolutions. Within general relativity, the method has recently been employed e.g. by \cite{KopacekKKS-10} in their study of orbital dynamics of a charged test particles around a rotating black hole immersed in a magnetic field.

A very useful tool for visualisation of the recurrences are {\it recurrence plots}, introduced by \cite{EckmannOR-87}. Suppose one knows the phase-space trajectory with constant step of time (we use proper time in our case of geodesic motion in a given space-time).\footnote
{The knowledge of just one phase variable (e.g. position) is enough in fact, since the phase space can be reconstructed from a sequence of its time-delayed series as in the directional-vectors method discussed in previous section.}
Denoting by $\vec{X}_i=\vec{X}(\tau_i)$ the $N$ successive points of the phase trajectory, one defines the so called recurrence matrix by
\begin{equation}
  R_{i,j}(\epsilon)=
  \Theta\left(\epsilon-\parallel\!\!\vec{X}_i-\vec{X}_j\!\!\parallel\right),
  \qquad i,j=1,...,N \;,
\end{equation}
where $\epsilon$ is the radius of a chosen neighbourhood (it is called threshold), $\parallel\cdot\parallel$ denotes the chosen norm (in accord with a common experience, the picture of long-term dynamics only slightly depends on which norm is chosen) and $\Theta$ is the Heaviside step function. The matrix thus contains only units and zeros and can be easily visualised by representing 1's by black dots (while 0's by blank spaces) at the respective coordinates $i$, $j$. For regular systems, the black points tend to arrange in distinct structures, in particular in long parallel diagonal lines (their distance scales with period) and checkerboard structures, whereas for random behaviours the black points are scattered without order. Chaotic systems yield the most ``artistic" plots: they contain blocks of almost-diagonal patterns as well as irregular ones, apparently placed one over another within horizontal and vertical structures. The main diagonal $R_{i,i}$ is trivial (``line of identity") and present in every system (for it is often omitted) and the matrix is symmetric with respect to it. The almost-regular blocks correspond to time intervals when the trajectory sticks to some unstable periodic orbit; the more unstable this orbit is, the earlier the trajectory deviates from it and so the smaller is the block. Horizontal/vertical lines indicate periods when the system is trapped in some region of phase space without much change.

Judging the prominence of diagonal or other patterns within the recurrence plot by pure observation is of course subjective, so several quantifiers of the recurrence-matrix properties have been proposed. The simplest of them is the recurrence rate, given by ratio of the recurrence points (black ones) within all points of the matrix,
\[RR(\epsilon)=\frac{1}{N^2}\sum_{i,j=1}^N R_{i,j}(\epsilon) \,.\]
Another, most important quantifier is the histogram of diagonal lines of a certain prescribed length $l$,
\begin{eqnarray*}
  P(\epsilon,l)
  &=& \sum_{i,j=1}^N [1-R_{i-1,j-1}(\epsilon)][1-R_{i+l,j+l}(\epsilon)]\times \\
  &~& \times \prod_{k=0}^{l-1}R_{i+k,j+k}(\epsilon) \,.
\end{eqnarray*}
Several further quantities can in turn be computed from this histogram. The one called $DET$ is given by ratio of the points which form a diagonal line longer than a certain $l_{\rm min}$ within all the recurrence points,
\[DET(\epsilon)=
  \frac{\sum_{l=l_{\rm min}}^N lP(\epsilon,l)}{\sum_{l=1}^N lP(\epsilon,l)} \;,\]
while the average length of diagonal lines is
\[L(\epsilon)=
  \frac{\sum_{l=l_{\rm min}}^N lP(\epsilon,l)}{\sum_{l=l_{\rm min}}^N P(\epsilon,l)}\;.\]
The length of the longest diagonal $L_{\rm max}(\epsilon)=\max_{i=1..N}\{l_i\}$ and its inverse $DIV(\epsilon)=1/ L_{\rm max}(\epsilon)$ are also of interest, since they are most directly related to the rate of divergence of nearby orbits and serve as a rough estimate of the largest Lyapunov exponent.

It is possible to make this estimate more precise, because the cumulative histogram of diagonals turned out to yield the second-order R\'enyi's entropy $K_2$ (also called correlation entropy) which stands for a lower estimate of the sum of positive Lyapunov exponents.
Let the phase space be divided in boxes with the size $\epsilon$, numbered in some order by $i_1,\dots,i_l$. Denote by $p_{i_1,\dots,i_l}(\epsilon)$ the probability that the point $\vec{X}(\Delta\tau)$ is in the $i_1$-th box, the following point $\vec{X}(2\Delta\tau)$ is in the $i_2$-th box, and so on, up to the point $\vec{X}(l\Delta\tau)$.
The second order R\'enyi's entropy $K_2$ is given by
\begin{equation}
  K_2(\epsilon,l)=-\lim_{\Delta\tau \to 0} \lim_{\epsilon \to 0} \lim_{l \to \infty}
                   \frac{1}{l\Delta\tau} \ln\sum_{i_1}^{i_l}p^2_{i_1,\dots,i_l}(\epsilon) \,.
\end{equation}
Realising that the occurrence of (non-trivial) diagonal of length $l$ means that $l$ successive points of the trajectory $\vec{X}(\tau)$, $\vec{X}(\tau+\Delta\tau)$, $\dots$, $\vec{X}(\tau+(l-1)\Delta\tau)$ lie in the $\epsilon$-neighbourhoods of certain other $l$ successive points $\vec{X}(\tau_0)$, $\vec{X}(\tau_0+\Delta\tau)$, $\dots$, $\vec{X}(\tau_0+(l-1)\Delta\tau)$, \cite{MarwanCTK-07} showed that (under the assumption of ergodicity) the sum can be approximated by
\begin{equation}
  \sum_{i_1}^{i_l}p^2_{i_1,\dots,i_l}(\epsilon)
  \approx
  \frac{1}{N}\sum_{t=1}^N p_t(\epsilon,l) \,,
\end{equation}
where $p_t(\epsilon,l)$ is the probability of finding the line of length $l$ in the boxes centered at points $\vec{X}(\tau)$, $\vec{X}(\tau+\Delta\tau)$, $\dots$, $\vec{X}(\tau+(l-1)\Delta\tau)$.
Hence, $K_2$ can be estimated from the relation
\begin{equation}
  K_2(\epsilon,l)\approx\hat{K}_2(\epsilon,l)=-\frac{1}{l\Delta\tau}\ln{p_c(\epsilon,l)}\,,
\end{equation}
where $p_c(\epsilon,l)$ is the probability of finding a diagonal line whose length is {\em at least} $l$. This means that $\hat{K}_2$ is determined by a slope of the cumulative histogram plotted (in logarithmic scale) against the diagonal length $l$. At the same time, the correlation entropy $K_2$ was shown to yield a lower estimate of the sum of positive Lyapunov exponents, so the estimate $\hat{K}_2$ is a good indicator of chaotic behaviour.

Similarly as with diagonal lines, one can plot the histogram of vertical (or horizontal) lines (against their length $v$),
\[P(\epsilon,v)=
  \sum_{i,j=1}^N (1-R_{i,j}(\epsilon))(1-R_{i,j+v}(\epsilon))
  \prod_{k=0}^{v-1}R_{i,j+k}(\epsilon) \,,\]
and also the respective measure of vertical structures (parallel of $DET$, called laminarity)
\[LAM(\epsilon)=\frac{\sum_{v=v_{\rm min}}^N vP(v)}{\sum_{v=1}^N vP(v)} \;.\]
The average length of vertical lines,
\[TT(\epsilon)=\frac{\sum_{v=l_{\rm min}}^N vP(v)}{\sum_{v=v_{\rm min}}^N P(v)} \;,\]
is called trapping time, because it indicates for how long the system is ``trapped" (without evolution in phase-space).
Finally, from the probabilities that some chosen diagonal or vertical line has length $l$, $p(l)=P(l)/N_l$ and $p(v)=P(v)/N_v$, where $N_l$ and $N_v$ are total numbers of diagonal/vertical lines, one can compute the so called Shannon entropies
\begin{eqnarray*}
  L\,ENTROPY&=&-\sum_{l=l_{\rm min}}^N p(l)\ln p(l) \,, \\
  V\,ENTROPY&=&-\sum_{v=v_{\rm min}}^N p(v)\ln p(v) \,.
\end{eqnarray*}

Also relevant as indicator is the size of the ``white gaps" between vertical (or horizontal) lines, because it is related to the recurrence times. For example, let us choose a point $\vec{X}_i$ on some trajectory and record the sequence of points which fall in its $\epsilon$-neighbourhood, $\{\vec{X}_{j_1},\vec{X}_{j_2},\dots\}$ (this set corresponds to black dots in a certain column/row of the recurrence matrix). Compute the recurrence times given by differences between serial numbers of the consecutive recurrence points $\vec{X}_{j_{k+1}}$, $\vec{X}_{j_k}$ multiplied by the respective proper-time steps, $T_k(\epsilon)=(j_{k+1}-j_k)\Delta\tau$. The mean of $T_k$ is called the recurrence time of the first type, $T1$. However, the recorded set usually also contains successive points between which the orbit does not leave the given $\epsilon$-neighbourhood, so $T_k(\epsilon)=1$. These points (called sojourn points) does {\em not} represent true recurrences, so they should be discarded from statistics. After removing all the sojourn points, the above recurrence set contains just beginnings of the vertical black lines. An average of their distances (average length of white vertical gaps) is called the 2nd-type recurrence time, $T2$. The recurrence times typically behave inversely to $RR$.

For the recurrence analysis to yield plausible results, it is crucial to set the parameters $\epsilon$, $\Delta\tau$, $l_{\rm min}$, $v_{\rm min}$ properly. The dependence of the results on these parameters is itself interesting to explore. In particular, the recurrence matrix and all the quantities computed from it depend critically on the ``target" size $\epsilon$. For example, when $\epsilon$ is chosen too large, the time step $\Delta\tau$ rather small and/or $l_{\rm min}$ (or $v_{\rm min}$) too small, there are plenty of sojourn points in the recurrence matrix. On the other hand, if $\epsilon$ is too small, the matrix may come out too sparse. The $l_{\rm min}$ limits the occurrence of short diagonal lines which are often formed, when $\epsilon$ is large enough, due to the fact that the $\epsilon$-neighbourhoods of the $n$-th and $(n+1)$-st recurrence loops have finitely long non-empty intersection, even though the loops may be quite diverging from each other. Finally, there is a subtle issue of misleading long secondary diagonals which was pointed out by \cite{MarwanCTK-07}; let us only touch on it here: an abundant occurrence of sojourn vertical sequences involving the main diagonal leads to the occurrence of another long diagonals in the main-diagonal vicinity, which deform the recurrence quantifiers if taken into account. Namely, the usage of $v_{\rm min}$ only excludes the short sojourn verticals from the statistics, but does not influence the statistics of diagonals; and the usage of $l_{\rm min}$ only excludes {\em short} diagonals, not the long ones. Therefore, in order to get rid of the latter, a certain lower bound for the length of vertical lines (called Theiler's parameter, $w$) is sometimes required {\em already in recording the recurrence matrix} (in addition to the careful choice of $\epsilon$), i.e. only the points are considered whose serial indices $i$ and $j$ satisfy $|j-i|\geq w$. We used the Theiler's parameter only in computing figures \ref{10=6-RPs} and \ref{K2-epsilon} (right column).

\begin{figure*}
\includegraphics[width=\textwidth]{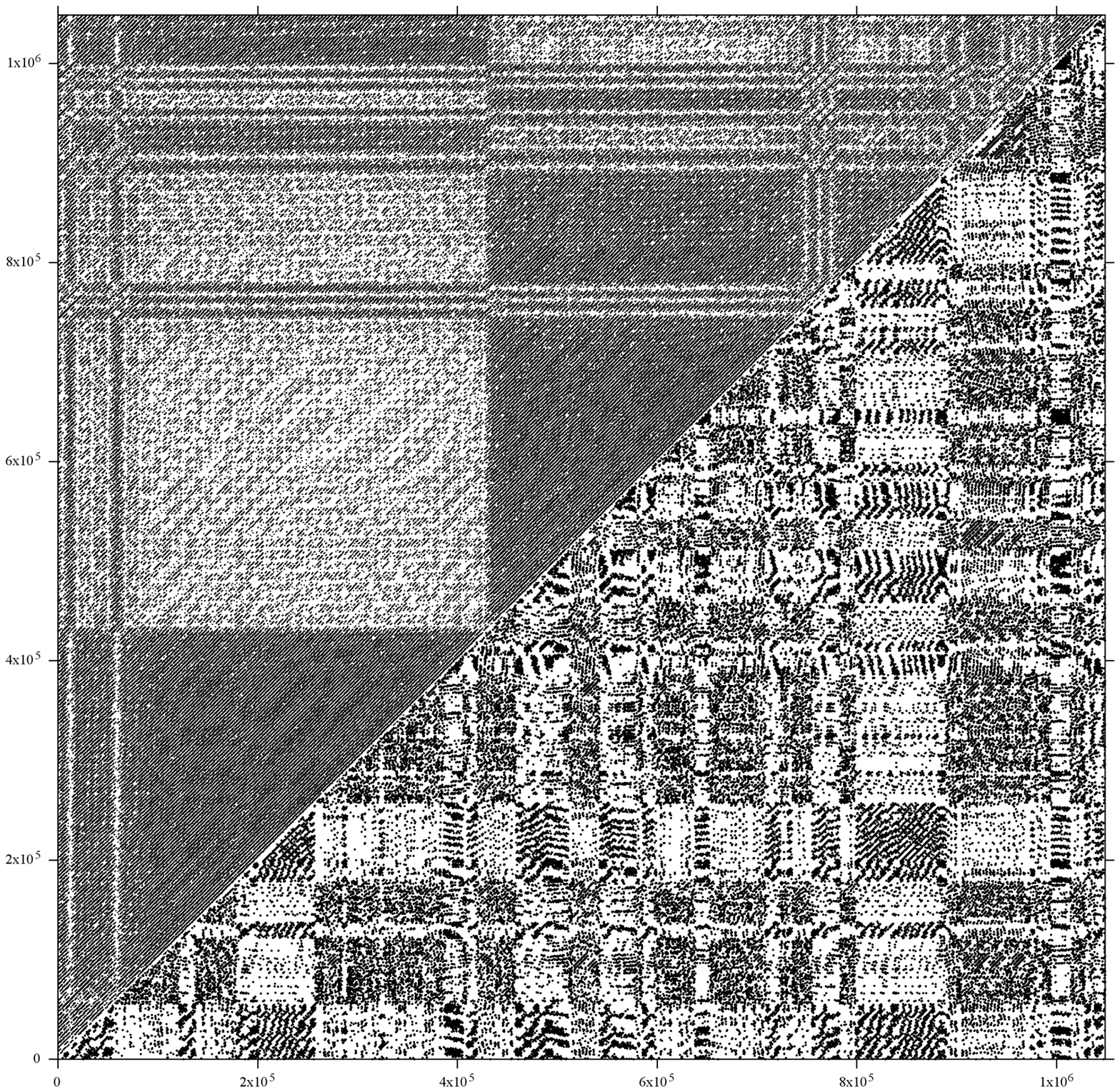}
\caption
{Recurrence plots for the two orbital sections whose Poincar\'e maps, $z(t)$ evolutions and their power spectra were shown in figure \ref{orbit1-excerpts}. We take advantage of the recurrence-matrix symmetry and give just halves of the recurrence plots in one square: the almost regular section (the left column of figure \ref{orbit1-excerpts}) is above the main diagonal (upper left triangle), while the rather chaotic section (the right column of figure \ref{orbit1-excerpts}) is below the main diagonal (lower right triangle). The axis values indicate proper time in units of $M$.}
\label{orbit1-excerpts-RPs}
\end{figure*}

\begin{figure*}
\includegraphics[height=657pt]{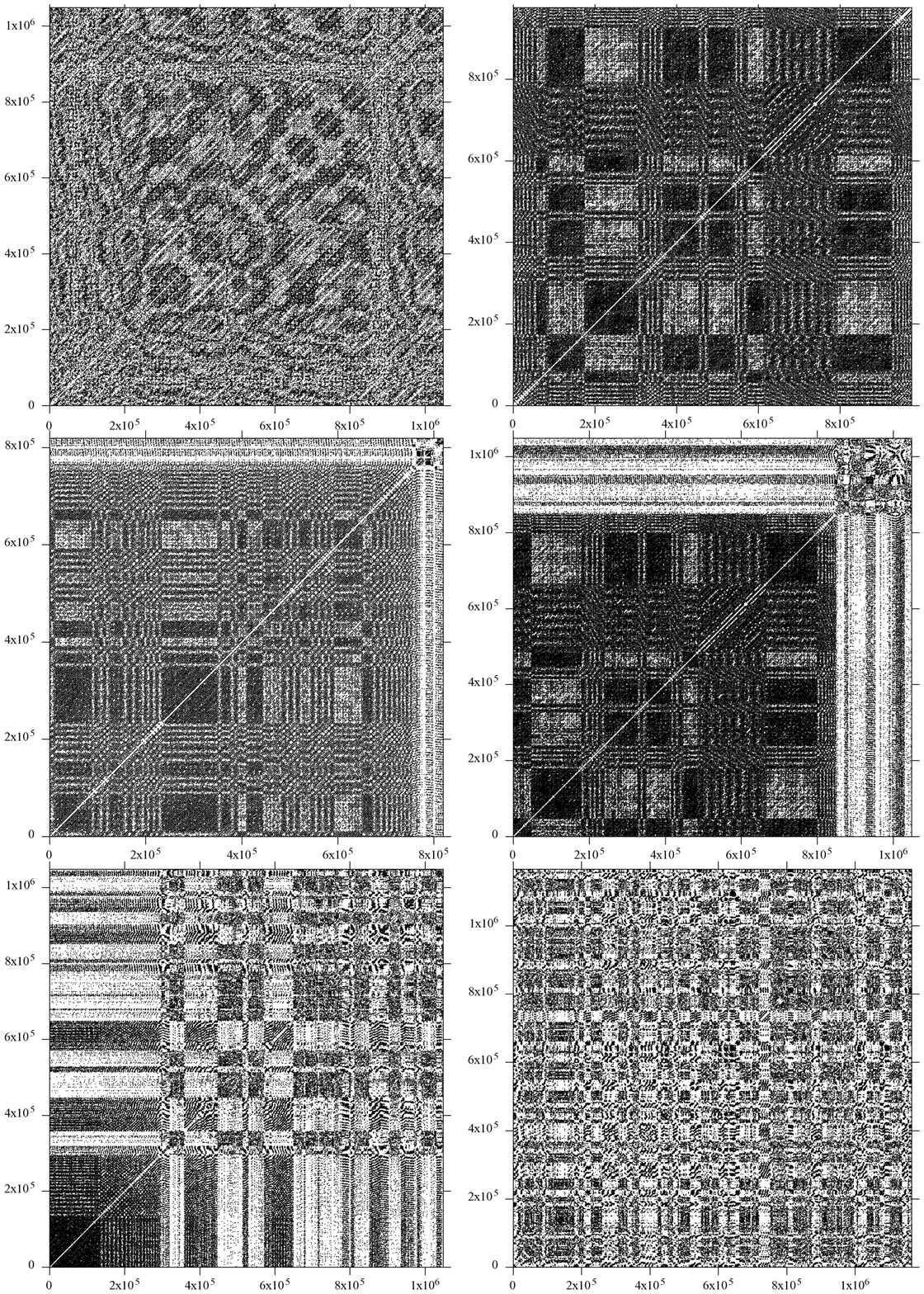}
\caption
{Recurrence plots for the orbital sections whose $z(t)$ power spectra and Kaplan-Glass parameter $\bar\Lambda(\Delta\tau)$ were shown in figures \ref{5-spectra} (A and B). The $z(t)$ evolutions and power spectra corresponding to the middle-row plots are given in the first two rows of figure \ref{series}. Proper time goes along the axes in units of $M$.}
\label{9=5-RPs}
\end{figure*}

\begin{figure*}
\includegraphics[width=\textwidth]{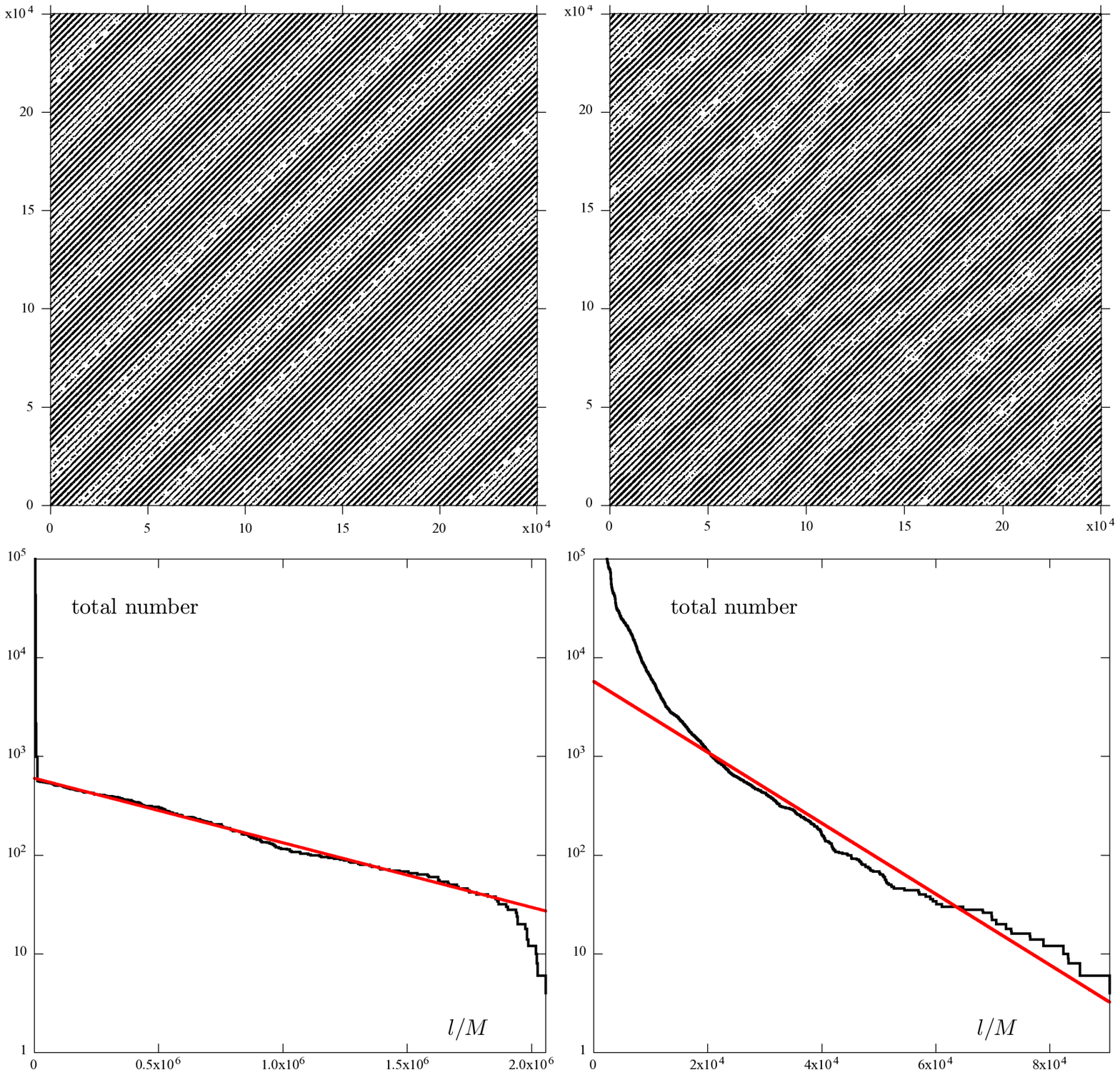}
\caption
{Recurrence plots and histograms of diagonal-line lengths for the two orbits whose Poincar\'e diagrams, $z(t)$ power spectra and tangent's autocorrelation parameter were shown in figure \ref{iMM-other}. Both orbits are rather regular: left orbit belongs to five-periodic regular islands and the right one sticks to them. Both have been followed up to $2.1\cdot 10^6\,M$ of proper time, with $15M$ step; we set $\epsilon=1.1$ and employed the Theiler's parameter $w=4$. The recurrence analysis clearly distinguishes between the orbits, in particular, the R\'enyi's entropy $K_2$, read off from the slope of the cumulative histogram plotted (in logarithmic scale) against the diagonal length $l$ (bottom plots), comes out about $-1.5\cdot 10^{-6}$ for the regular orbit while about $-8.27\cdot 10^{-5}$ for the ``sticky" orbit (notice that the $x$-axis ranges are different, so the difference in slopes is much bigger than how it appears at first sight). Obviously the slope has to be determined from the middle part of the histogram which really reflects recurrence properties and where it can be approximated by a straight line.}
\label{10=6-RPs}
\end{figure*}

\begin{figure*}
\includegraphics[width=\textwidth]{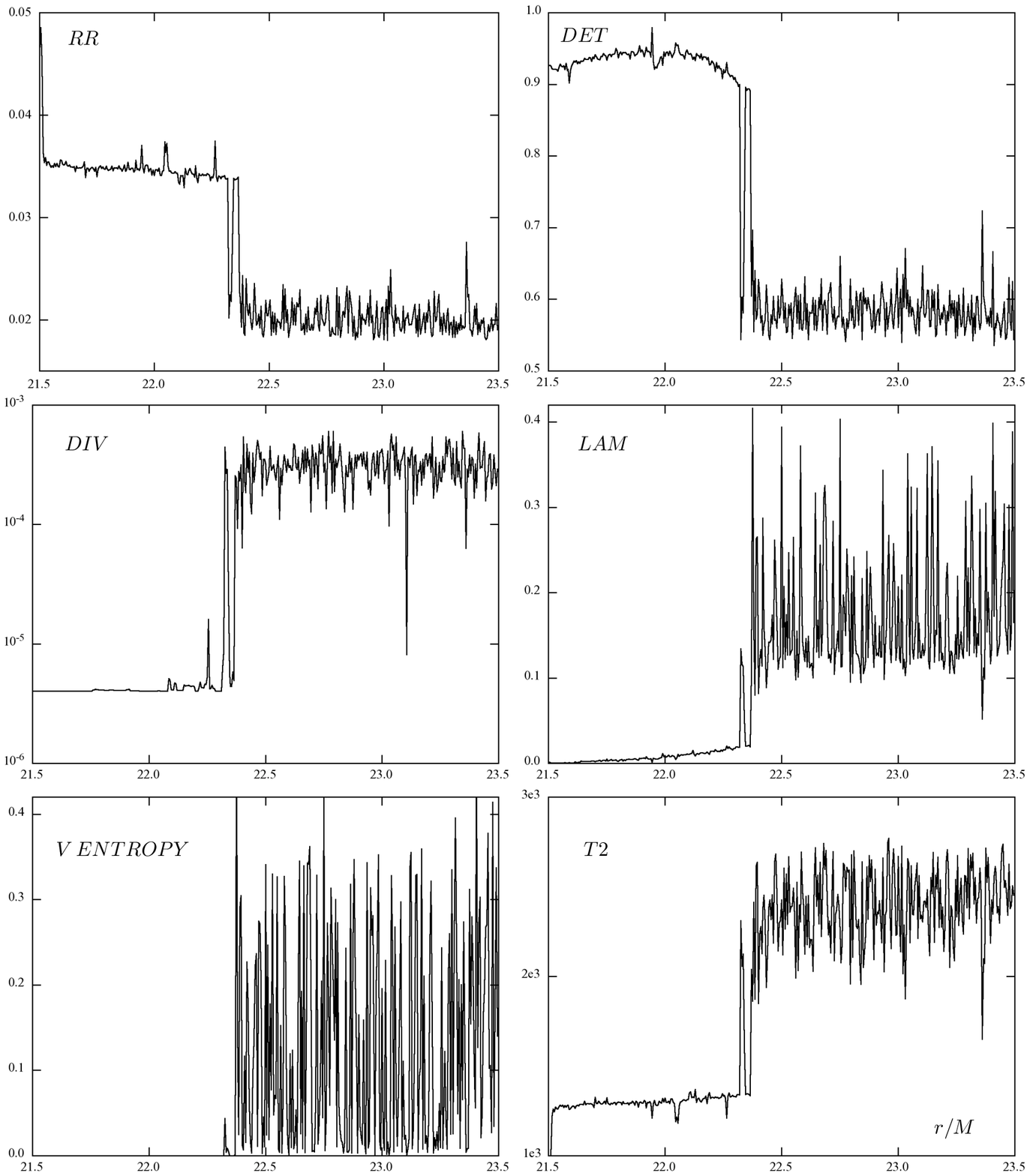}
\caption
{Six of the recurrence-analysis quantifiers, computed for 400 geodesics ejected tangentially (with $u^r=0$) from radii between $r=21.5M$ and $r=23.5M$ (with step $0.005M$) from the equatorial plane of the system of a black hole (of mass $M$) and the inverted 1st Morgan-Morgan disc with mass $1.3M$ and inner Schwarzschild radius $20M$. All the orbits have specific energy ${\cal E}=0.934$ and specific angular momentum $\ell=4M$. The orbits have been followed for about $250000M$ of proper time (thus the minimal achievable $DIV$ is $4\cdot 10^{-6}$) with ``sampling period" $\Delta\tau=45M$, the minimal length of diagonal/vertical lines has been set at $90M$ and the recurrence threshold at $\epsilon=1.1$. The left part of the plots scans trajectories belonging to a large secondary regular island, while the right part goes through the chaotic sea. All the quantifiers clearly distinguish between the two regimes.}
\label{RQA-a}
\end{figure*}

\begin{figure*}
\includegraphics[width=\textwidth]{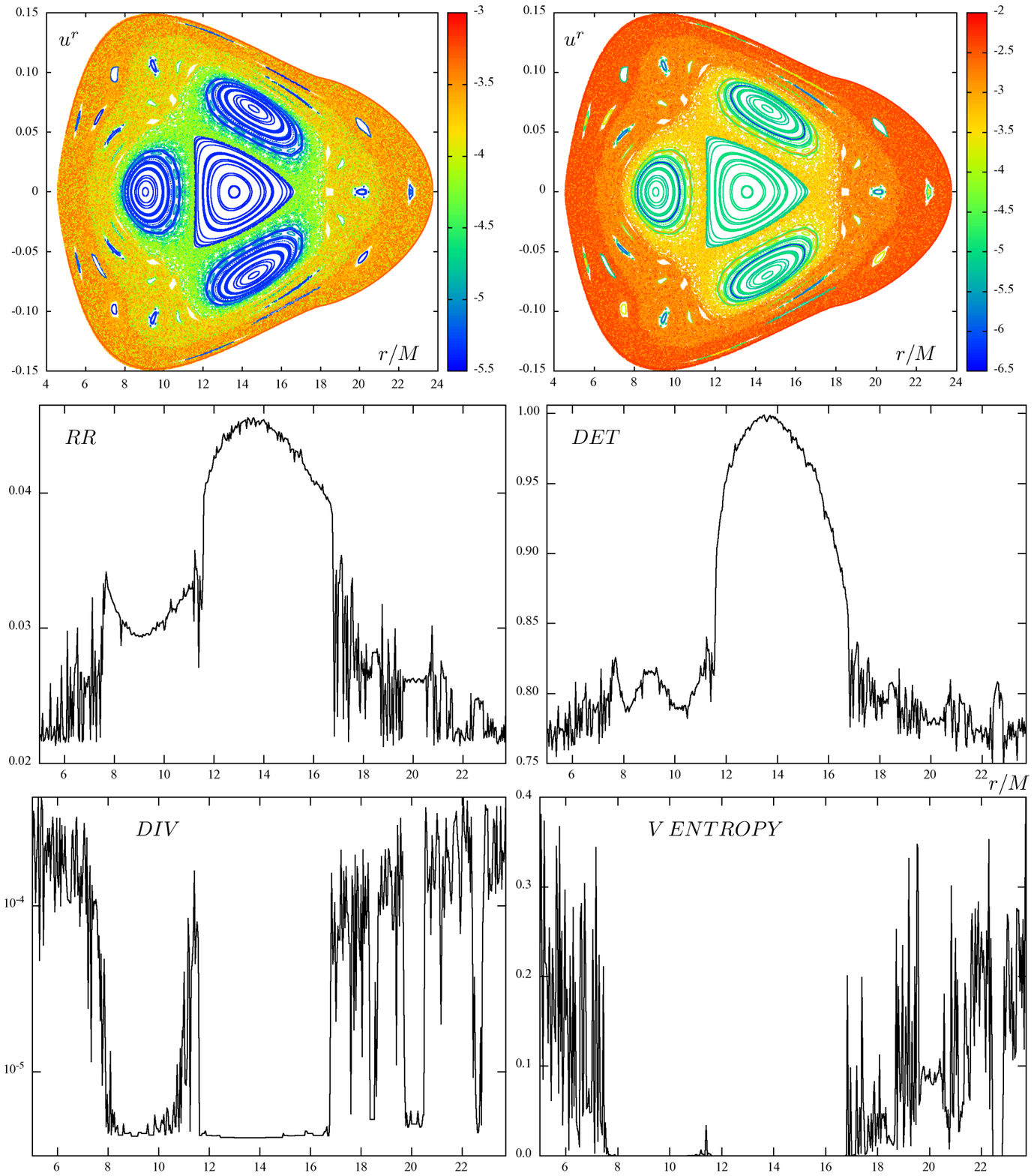}
\caption
{The quantifiers $RR$, $DET$, $DIV$ and $V\,ENTROPY$ computed, like in figure \ref{RQA-a}, for 470 geodesics ejected tangentially (with $u^r=0$) from radii between $r=5M$ and $r=24M$ (by $0.04M$) from the equatorial plane of the system of a black hole ($M$) and the inverted 1st Morgan-Morgan disc with mass ${\cal M}=0.5M$ and inner radius $r_{\rm disc}=18M$. All the orbits have specific energy ${\cal E}=0.9532$ and specific angular momentum $\ell=3.75M$. The orbits have again been followed for about $250000M$ of proper time with ``sampling period" $\Delta\tau=45M$, the minimal length of diagonal/vertical lines has been set at $90M$ and the recurrence threshold at $\epsilon=1.25$. Poincar\'e diagram of the system is shown at the top, with orbits coloured according to the value of their $DIV$ (top left) and according to the slope of the diagonal-line histogram (top right); the scale is logarithmic. A system with a more complicated phase portrait has been chosen here in order to compare the results with the simple case of figure \ref{RQA-a} and in order to check how sensitive the quantifiers are to more tiny phase-space features.}
\label{RQA-b}
\end{figure*}

\begin{figure*}
\includegraphics[width=\textwidth]{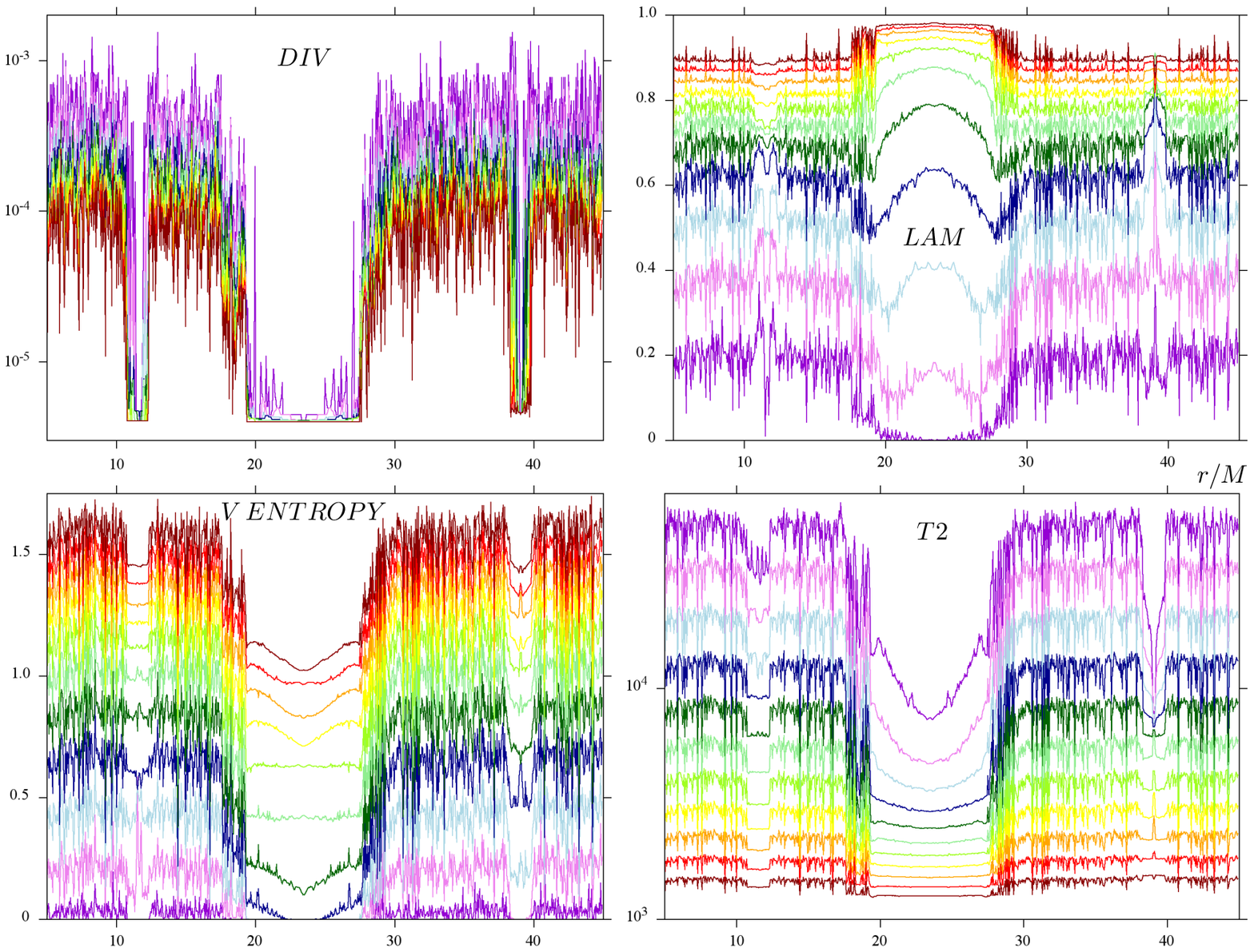}
\caption
{Dependence of four recurrence quantifiers on the choice of the threshold approach $\epsilon$. We go through the geodesics starting with $u^r=0$ from the radii $r=5M\div 45M$ from the equatorial plane of the black-hole--disc system with the disc mass ${\cal M}=1.3M$ and inner disc radius $r_{\rm disc}=20M$. All the geodesics have specific energy ${\cal E}=0.956$ and specific angular momentum $\ell=4M$. Their $DIV$, $LAM$, $V\,ENTROPY$ and $T2$ parameters are computed for 11 different values of $\epsilon$, namely $0.50$, $0.65$, $0.80$, $0.95$, \dots $1.85$, $2.00$; in this order, the line colour shifts from dark violet, violet, light blue, dark blue, \dots, to red and dark red. A primary regular island and two smaller islands aside are evident.}
\label{RQA-epsilon}
\end{figure*}

\begin{figure*}
\includegraphics[width=\textwidth]{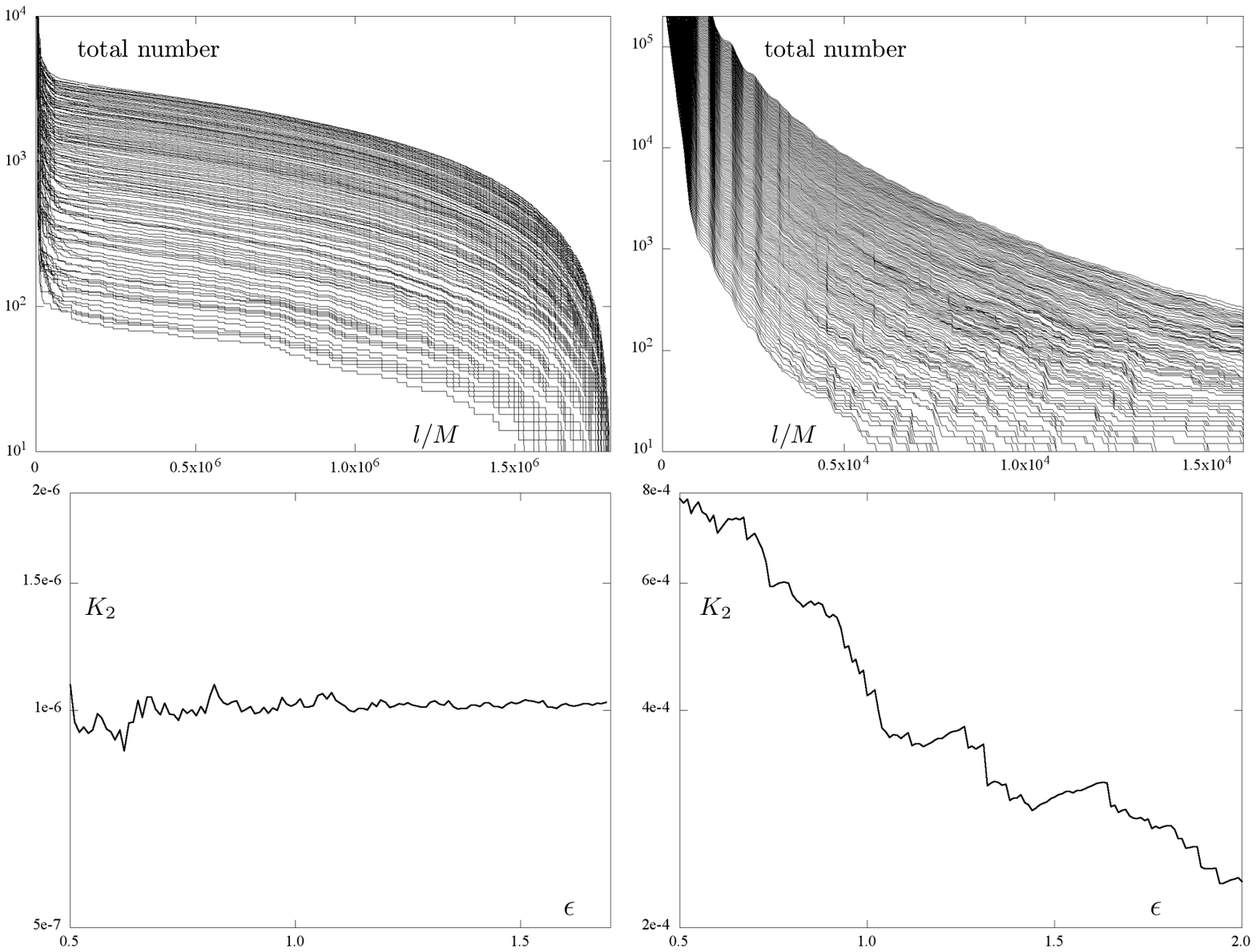}
\caption
{Dependence of the diagonal-line cumulative length-histogram (top row) and of the resulting value of the correlation entropy $K_2$ (bottom row) on the choice of the threshold approach $\epsilon$, plotted for a regular (left column) and weakly chaotic (right column) geodesics. Both geodesics have been followed for about $2\cdot 10^6M$ of proper time. Both have specific energy ${\cal E}=0.956$ and specific angular momentum $\ell=4M$ and live in a space-time of a black-hole ($M$) surrounded by the inverted 1st Morgan-Morgan disc with mass ${\cal M}=1.3M$ and inner radius $r_{\rm disc}=20M$. The recurrence matrix of the chaotic geodesic has been computed using the Theiler parameter growing linearly with $\epsilon$ (from $w\!=\!2$ to $w\!=\!6$). The $K_2$ entropy of this orbit is considerably more sensitive to $\epsilon$ than that of the regular orbit on the left (the range along both $K_2$-axes corresponds to a change by a factor of 4), however the two cases (regular/chaotic) are clearly distinguished as seen on the orders at the vertical axes.}
\label{K2-epsilon}
\end{figure*}

Several examples of recurrence plots obtained for our black-hole--disc system are attached.
For computation of the recurrence matrix, we used the Euclidean norm (both position and velocity are processed as vectors in Euclidean space) and ``Cartesian" spatial mesh corresponding to spherical Schwarzschild coordinates; each of the three coordinates has been normalised to zero mean value and unit standard deviation.
First, we drew the recurrence plots for orbits (or their parts) studied in previous sections in order to compare the information they reveal with that provided by Poincar\'e maps, time series and their spectra, and by the average-directional-vector method. Figure \ref{orbit1-excerpts-RPs} shows recurrence plots of the two orbital sections (one rather regular and one rather chaotic) compared in figure \ref{orbit1-excerpts}. The almost regular section occupies the upper left half and the rather chaotic section occupies the lower right half; the contrast in their ``entropy" is obvious. Figure \ref{9=5-RPs} brings recurrence plots of the six orbital sections whose $z=z(t)$ power spectra and Kaplan-Glass averaged autocorrelation parameter $\bar\Lambda(\Delta\tau)$ were presented in figures \ref{5-spectra}.A and \ref{5-spectra}.B, respectively. The plots are again ordered in the direction of increasing chaoticity. The first of them really looks like one of the dark boxes of the subsequent plots (these boxes correspond to ``sticky-motion" periods when the orbit is tied to a regular region); the second plot shows a clear checkerboard pattern (the orbital section is rather regular) which then gradually give way to horizontal/vertical pattern at times when the orbit's behaviour changes to chaotic.\footnote
{Note that the middle-row plots of this figure are counterparts of the $z(t)$ time series and spectra shown in the first two rows of figure \ref{series}. One can check by comparison that the transition to chaos in the $z(t)$ evolution really occurs at times when the recurrence matrix changes from checkerboard to horizontal/vertical regime (``carpet edge" in the recurrence plots).}
Figure \ref{10=6-RPs} brings recurrence plots and diagonal-length histograms for the two rather regular orbits whose Poincar\'e diagrams, $z(t)$ power spectra and tangent's autocorrelation parameter were shown in figure \ref{iMM-other}. The recurrence analysis well distinguishes between the orbits, in particular, the histogram slope (indicated in red in the plots) which yields R\'enyi's entropy is steeper for the less regular orbit.

Figures \ref{RQA-a} and \ref{RQA-b} show six (the latter only four) of the recurrence-analysis quantifiers ($RR$, $DET$, $DIV$, $LAM$, $V\,ENTROPY$ and $T2$), computed for large collections of orbits ejected from certain radial ranges within the equatorial plane of the black-hole--disc systems. The quantifiers evidently distinguish between regular and chaotic orbits, but their sensitivity to different phase-space features is somewhat different. In the figure \ref{RQA-a} the difference between regular and chaotic region is more distinct, because the orbits in the left part belong to a large regular island which then quite immediately passes into a chaotic sea (right part of the plots). On the other hand, the figure \ref{RQA-b} scans a more complex phase-space region where several regular islands and chaotic layers are crossed. For a better picture of which region have been considered, the respective Poincar\'e diagram is also shown in figure \ref{RQA-b}. In the chaotic regime, the recurrence rate and the fraction of diagonal lines are seen to be lower, whereas the vertical measures $T2$ and $V\,ENTROPY$ grow significantly. This means that the system takes more time to get back to the chosen $\epsilon$-neighbourhood and that the vertical-line length oscillates more than in the regular regime; namely, in the regular regime, there are either no vertical lines at all, or they are of approximately the same length (note, for example, that if $v_{\rm min}$ is chosen too small, one gets a lot of ``sojourn" vertical lines of the same length). The $DIV$ parameter of course increases in the chaotic region as well as the ``laminarity". All the quantifiers get considerably more noisy in the chaotic regions.

In order to illustrate how Poincar\'e diagrams can be supplemented by information provided by recurrence analysis, the passage points are coloured according to the values of selected recurrence quantifiers which apply to the respective orbits (so the colours are of course mixed in chaotic regions); namely, the top left diagram is coloured according to the value of $DIV$ and the top right diagram according to the slope of the diagonal-line histogram. We have chosen exactly these two quantifiers, because $DIV$ is one of the simplest, whereas the diagonal-histogram slope is more ``sophisticated", tightly connected with Lyapunov exponents. However, it is seen that the two colourings provide almost the same information (only the colour scales are somewhat different, naturally). Therefore, the $DIV$ quantifier seems to be more suitable for a quick distinction between regularity and chaos, mainly if large sets of trajectories are to be processed routinely. The cumulative-histogram slope {\em is} theoretically more sophisticated, but also harder to get (it is ``higher-level") and, mainly, its ``automatic" computer evaluation is much more tricky, because it has to be determined from a proper part of the histogram. Namely, only a certain middle part of the histogram is relevant, since its short-length end typically ``diverges" due to increasing number of sojourn points, while the long-length end typically falls off quickly due to the finite length of the trajectory. (The histogram has to be computed in the limit $l\to\infty$, so the short-length end has actually no sense; the long-length end is of course determined by the fact that practically the trajectories cannot be infinitely long.)

Finally, the dependence of several quantifiers on the choice of $\epsilon$ is illustrated in figure \ref{RQA-epsilon}. There, the behaviour of $DIV$, $LAM$, $V\,ENTROPY$ and $T2$ over geodesics starting from a wide range of radii is plotted for 11 different values of $\epsilon$. In all the plots, one can clearly recognise a large primary regular island and two smaller ones. The quantifiers generally show monotonous dependence on $\epsilon$ --- $LAM$ and $V\,ENTROPY$ increase whereas $DIV$ and $T2$ decrease with $\epsilon$, as expected.
Figure \ref{RQA-epsilon} may also serve as a further support for the $DIV$ quantifier (see figure \ref{RQA-b} as well). As also confirmed by other similar figures we are not showing here, the large regular islands can be recognised easily by any of the quantifiers, because the latter fluctuate there much less and around markedly different values than in the chaotic regions. But only the $DIV$ parameter appears to be fairly indicative of small islands as well. Namely, in regular regions (both large and small) it is typically 2 orders of magnitude lower than in chaotic regions ($DIV\sim 3\times 10^{-6}$ within regular regions, about an order higher in thin chaotic layers and about $3\times 10^{-4}$ in large chaotic regions; this chaotic-sea value corresponds to a divergence time of several thousands $M$, which represents some ten cycles about the black hole). For longer trajectories (than those we have treated here) the difference in $DIV$ between regular and chaotic regions tends to be even bigger. The other quantifiers only respond to small islands by reducing their oscillation, but not by a noticeable change of the mean value. Loosely speaking, their sensitivity profile is shifted towards larger regular regions (see the $RR$ quantifier in figure \ref{RQA-b}, for example). The last figure \ref{K2-epsilon} illustrates the difference between regular and chaotic geodesic on a markedly different value of $K_2$ entropy inferred from the cumulative length-histogram of diagonal lines, and also on different dependence on $\epsilon$ of this histogram.

\section{Concluding remarks}

In paper I, we performed an overall check of the time-like geodesic dynamics in the field of a Schwarzschild black hole surrounded by an axially symmetric static thin disc or ring, and of the dependence of its chaoticity on parameters characterising the additional source and the test particles. In the present paper, we have focused on individual orbits and their different parts and showed how the degree of their irregularity, already visible on Poincar\'e diagrams, can be judged in more detail by studying the time series obtained for phase variables. We have mainly considered the times series $z(t)$ of the position perpendicular to the disc/ring plane, computed their power spectra and compared the information thus gained with the one provided by the method of Kaplan \& Glass which tracks autocorrelation between different parts of the series in dependence on time shift, and also with outcomes of the recurrence-matrix analysis of Eckmann et al. All these methods prove simple and powerful, while they differ in sensitivity to specific types of behaviour.

Let us also include a brief mention of several others' results which appeared in the field recently.
\cite{Brink-08} contributed to the topic of complete geodesic integrability, mentioned in the introduction, by studying geodesics in stationary axisymmetric vacuum space-times and the correlation of its fabric with the existence of the ``fourth integral". (Cf. also \cite{Markakis-12} for a Newtonian treatment of the integrability problem.)
\cite{VerhaarenH-10} returned to the study of dynamics of test particles with spin in a Schwarzschild space-time and argued, on Poincar\'e diagrams and Lyapunov exponents, that smaller values of spin than previously thought can already make the particle motion chaotic.
\cite{KovacsBT-11} performed the first post-Newtonian analysis of the Sitnikov system (motion of a test body in the field of an equal-mass binary, along a line perpendicular to the orbital plane and going through the barycentre) and demonstrated, by numerical study of the system in dependence on gravitational radius of the ``primaries", that the relativistic effect of pericentre advance does not destroy its chaotic aspects.
\cite{Ramos-CaroPL-11} studied the motion of test particles in the field of a centre with quadrupole deformation surrounded by finite thin discs obtained by superpositions of members of a counter-rotating Morgan-Morgan family. They found there is a close connection between linear stability/instability of equatorial circular orbits and regularity/chaoticity of general three-dimensional orbits passing through their radii.\footnote
{We would like to remember professor Patricio Letelier who was a leading expert in the fields of general relativity and chaotic dynamics. He left us just at the time when the above paper appeared in MNRAS.}
The same group \citep{LetelierRL-11} also revisited geodesic dynamics in the system composed of a monopole or an isotropic harmonic oscillator and oblate quadrupole, and found several new features not noticed before.
\cite{WangW-11} employed a pseudo-Newtonian potential in order to superpose a rotating black hole with a quadrupole halo. They analysed emission of gravitational waves from particles orbiting in such a field and demonstrated that the radiated amplitude and power are closely related to the degree of chaoticity of the orbit.
\cite{Galaviz-11} considered the evolution of a compact binary perturbed by a third body. Within a post-Newtonian version of the Hamiltonian ADM formulation, and using basin-boundary analysis and Lyapunov exponents, he examined the relative importance of different PN orders in inducing chaos in the system.
Very recently, \cite{ContopoulosHL-12} have analysed the classical system of two coupled oscillators and free motion in the general relativistic Manko-Novikov space-time (which describes a rotating axisymmetric compact body); they mainly focused on periodic orbits of the systems and on dependence of their properties on orbital energy.
Finally, \cite{Igata-etal-11} observed on Poincar\'e maps that the time-like geodesics bound in the field of the Emparan-Real 5D black ring show chaotic features.

To conclude, it should be admitted that once the evolution of a given dynamical system is mastered with sufficient numerical accuracy, it is rather easy to produce various decorative figures. But it is also true that these can indeed yield a good picture of how much irregular the system is and how the irregularity depends on system parameters. This in turn indicates how much the ``perturbations" acting on real systems degrade the corresponding simplified exact models and helps to evaluate the validity of various approximations. Turning to our particular problem of orbiting in a static compact-centre space-time, it would be suitable to employ such methods in order to estimate and compare the significance of various ``perturbations" present in real astrophysical situations. Besides the gravitational influence of additional matter, discussed (in a simple, static and axially symmetric case) in the present work, there would also occur {\em mechanical} interaction of the orbiter with that matter (see e.g. \citealt{SubrK-05}); both the compact centre and the matter around would probably have non-zero angular momentum, so dragging effects should be incorporated; actually the orbiter may itself be endowed with spin or even higher moments; incoming gravitational waves can perturb the system as well as those emitted by the orbiter (back reaction); of course, if the orbiter was charged, it would also be affected by electromagnetic field, if there is some around.

When thinking about possible perturbations of the originally completely integrable problem of free test motion in a Schwarzschild or Kerr field, one has mainly in mind the motion of individual stars around supermassive black holes in galactic nuclei. As already stressed in Introduction, there are in fact whole {\em clusters} of stars in galactic nuclei, so when solving the motion of an individual satellite, one should also take into account gravity of the whole cluster, or solve that motion right as a part of the problem of $N$ interacting bodies. Such a problem is difficult within general relativity, mainly if the black-hole centre and the disc or ring/toroid should also be taken into account, but it is being considered within Newtonian theory (see e.g. \citealt{Subr-etal-04}; \citealt{Haas-etal-11a}; \citealt{Haas-etal-11b}, and references therein) as well as in post-Newtonian and post-Minkowskian approximation (e.g. \citealt{Chu-09}; \citealt{LedvinkaSB-08}; \citealt{HartungS-11}).

There are several immediate options for further work. One can of course check the results with yet other methods (Lyapunov-type coefficients, various other ``entropies", basin-boundary analysis, Melnikov integral, etc.), find and study particular significant orbits of the system (periodic and homoclinic/heteroclinic orbits) in detail, or compare the relativistic analysis with Newtonian, pseudo-Newtonian or post-Newtonian one. However, we would mainly like to focus on astrophysically more realistic situations (parameter ranges) in the following. This should involve, among others, the question of whether to incorporate also another gravitating components like spheroidal halo, disc {\em plus} outer ring (or thick toroid), or/and jets, and---most importantly---the question of how to account {\em adequately} for rotation.

\section*{Acknowledgements}

We are grateful to M. \v{Z}\'a\v{c}ek who provided us the numerical code developed within his diploma work on geodesic motion in Weyl space-times.
The plots were produced with the help of the Gnuplot utility (we thank K. Houfek for hints) and D. Krause's {\tt bmeps} program.
For the recurrence analysis we were using the ``commandline recurrence plots" software (version 1.13z) by N. Marwan; we also thank O. Kop\'a\v{c}ek for discussions on the recurrence method.
Our work has been supported from the projects GACR-202/09/0772 and MSM0021620860 (O.S.); GACR-205/09/H033, SVV-265301 and GAUK-428011 (P.S.).

\end{document}